\documentclass[twocolumn]{revtex4}
\usepackage{graphicx}
\usepackage{dcolumn}
\usepackage{amsmath}
\usepackage{longtable}

\newcommand{\nbar}{\ensuremath{\langle N \rangle}}
\newcommand{\wn}{\textrm{cm}\ensuremath{^{-1}}}
\newcommand{\etal}{\emph{et al.}}
\newcommand{\vi}[1]{\ensuremath{\nu_{#1}}}
\newcommand{\pt}[2]{\ensuremath{#1\times 10^{#2}}}
\newcommand{\oH}{oH$_2$}
\newcommand{\pH}{pH$_2$}
\newcommand{\oD}{oD$_2$}
\newcommand{\pD}{pD$_2$}

\newcommand{\si}[1]{\ensuremath{_{\textrm{#1}}}}

\newcommand{\micro}{\ensuremath{\mu}}

\begin{document}

\preprint{JCP review}

\title{Helium nanodroplet isolation ro-vibrational spectroscopy:
  methods and recent results}

\author{Carlo Callegari}
\author{Kevin K. Lehmann}
 \email{lehmann@princeton.edu}
\author{Roman Schmied}
\author{Giacinto Scoles}
 \email{gscoles@princeton.edu}
\affiliation{Department of Chemistry, Princeton University, Princeton,
  New Jersey 08544}

\date{\today}

\begin{abstract}
%\vspace{4mm}
  In this article, recent developments in HElium NanoDroplet Isolation
  (HENDI) spectroscopy are reviewed, with an emphasis on the infrared
  region of the spectrum. We discuss how molecular beam spectroscopy
  and matrix isolation spectroscopy can be usefully combined into a
  method that provides a unique tool to tackle physical and chemical
  problems which had been outside our experimental possibilities.
  Next, in reviewing the experimental methodology, we present design
  criteria for droplet beam formation and its seeding with the
  chromophore(s) of interest, followed by a discussion of the merits
  and shortcomings of radiation sources currently used in this type of
  spectroscopy. In a second, more conceptual part of the review, we
  discuss several HENDI issues which are understood by the community
  to a varied level of depth and precision. In this context, we show
  first how a superfluid helium cluster adopts the symmetry of the
  molecule or complex seeded in it and discuss the nature of the
  potential well (and its isotropy) that acts on a solute inside a
  droplet, and of the energy levels that arise because of this
  confinement. Second, we treat the question of the homogeneous versus
  inhomogeneous broadening of the spectral profiles, moving after this
  to a discussion of the rotational dynamics of the molecules and of
  the surrounding superfluid medium.  The change in rotational
  constants from their gas phase values, and their dependence on the
  angular velocity and vibrational quantum number are discussed.
  Finally, the spectral shifts generated by this very gentle matrix
  are analyzed and shown to be small because of a cancellation between
  the opposing action of the attractive and repulsive parts of the
  potential of interaction between molecules and their solvent. The
  review concludes with a discussion of three recent applications to
  (a) the synthesis of far-from-equilibrium molecular aggregates that
  could hardly be prepared in any other way, (b) the study of the
  influence of a simple and rather homogeneous solvent on large
  amplitude molecular motions, and (c) the study of mixed
  $^{3}$He/$^{4}$He and other highly quantum clusters (e.g., H$_2$
  clusters) prepared inside helium droplets and interrogated by
  measuring the IR spectra of molecules embedded in them. In spite of
  the many open questions, we hope to convince the reader that HENDI
  has a great potential for the solution of several problems in modern
  chemistry and condensed matter physics, and that, even more
  interestingly, this unusual environment has the potential to
  generate new sets of issues which were not in our minds before its
  introduction.
\end{abstract}

\pacs{PACS number}% PACS, the Physics and Astronomy Classification Scheme.
\maketitle

%\tableofcontents

\section{Introduction}
\label{sec:intro}

\subsection{Scope of the review}
\label{sec:scope}

In this article we review the motivations and the experimental
methodology of HElium NanoDroplet Isolation (HENDI) spectroscopy, with
an emphasis on the infrared region of the spectrum. After reviewing
spectrometer design criteria and experimental know-how, we summarize
our present understanding of line spacings, line shapes and spectral
shifts, and discuss some of the recent results in as far as they
illustrate general principles and possible future applications.  Our
aim is centered on discussing the current understanding of the field,
on identifying unsolved problems and open questions, and not on
providing a complete summary of the work published so far, because the
last three years have seen a really explosive growth in this area.
%%NEW
For earlier reviews of this field, see
Refs.~\onlinecite{Toennies98,Whaley98}.
Before describing the motivations and general goals of this kind of
research, we shall first show how the natural evolution of supersonic
molecular beam infrared spectroscopy has led seamlessly to this
apparently esoteric, but in reality simple, extension, which, as we
hope to show here, is likely to spawn a large number of scientific and
technical applications.

\subsection{The marriage between molecular beams and matrix
  spectroscopy}
\label{sec:molbeams}

Because of the long fluorescence lifetimes of vibrationally excited
molecules, it is impractical, with few
exceptions~\cite{Stewart83,Chang93}, to obtain infrared spectra in
free jets using fluorescence detection. At the same time, when the use
of well collimated beams is called for, direct measurement of the
attenuation of the exciting radiation is also out of the
question~\cite{Miller92}.

This leaves bolometric detection of the energy deposited in the beam
by a resonant, tunable laser as the method of choice for this type of
spectroscopy~\cite{Gough77}. Assuming unity partition function and
full saturation of the IR transition, it can be shown that with a
seeded beam of ``standard'' intensity and the typical sensitivity of a
doped Si (or Ge) bolometer operating at 1.5~K (noise equivalent power
on the order of 100~fW/$\sqrt{\textrm{Hz}}$), signal-to-noise (S/N)
ratios of up to 10$^{5}$ can be obtained~\cite{Miller92}.

The method is equally applicable to van der Waals complexes which,
again with very few exceptions~\cite{Nesbitt94}, dissociate upon
vibrational excitation. What is detected in this case is a
\textit{decrease} of the total kinetic energy carried by the beam,
because the dissociation fragments are deflected from the beam axis,
resulting in a decreased particle flux to the detector.

If the complex consists of a large number of monomer units, it is not
\textit{a priori} obvious that absorption of energy must lead to
evaporation, as the amount of excess energy per degree of freedom can
conceivably be too small. One must consider, however, that the
temperature of large van der Waals clusters is determined by
evaporative cooling~\cite{Klots85} which is a self-limiting process.
It is then easy to appreciate that the absorption of an IR photon by a
large van der Waals cluster will in general put the cluster above the
threshold for evaporation to resume.
%For a beam of molecules seeded in
%a helium jet, the forward kinetic energy per molecule is of the same
%magnitude as that of a typical IR photon, hence the S/N ratios
%observed in photodissociation spectroscopy are similar to those of
%isolated molecules.

Using this detection technique and concepts, IR spectra of hundreds of
van der Waals complexes (mostly below 4 monomer units) have been
obtained during the last 20 years, contributing a great deal to our
knowledge of intermolecular forces~\cite{Hutson90,Nesbitt94}.  For
larger sizes, many experiments investigate a single chromophore
molecule complexed to a cluster of rare gas atoms, in order to keep
the interactions as simple as possible. This class of experiments has,
however, about an order of magnitude lower S/N ratios than the
experiments carried out on a seeded beam of stable molecules.

% Assuming a flux of 10$^{10}$~clusters/s (each of \mbox{$\approx 10^{3}$} Ar
% atoms in size, and each containing a single chromophore molecule)
% impinging on the detector in a set up analogous to that described
% above, a 3000~\wn photon will produce the evaporation of 5~Ar atoms
% per cluster, and therefore a power modulation on the detector of about
% 1~nW. Comparing with the numbers given above for energy-deposition
% spectroscopy of stable molecules, it follows that photoevaporation
% cluster spectroscopy is about a factor of 10 more difficult, mainly
% due to the lower number of chromophores present in the beam under
% typical conditions.

%\subsection{Matrix spectroscopy in the gas phase: helium nanodroplets
%  are an almost ideal spectroscopic matrix}
%\label{sec:clusterbeams}

In spite of the added difficulties, the possibility of loading a large
rare-gas host cluster with almost any chromophore by pickup of the
latter in a ``scattering'' cell located on the path of the host
cluster beam~\cite{Gough85} has generated a whole new class of
experiments.  As the exact number of atoms in the cluster starts
becoming inconsequential, this approach can be seen as the gas-phase
analog of Matrix Isolation (MI) spectroscopy~\cite{Gough83}. As in
bulk MI spectroscopy, several classes of experiments can be carried
out that involve both the unperturbed chromophore and its
photodissociation/reaction products.

In comparison with bulk MI spectroscopy, its Cluster Isolation (CI)
analogue has two main advantages and two major disadvantages.  The
first advantage is due to the finite size of the cluster matrix, which
allows control over the degree of chromophore complexation.
Uncontrolled aggregation upon annealing plagues MI spectroscopy and
sets one of the inevitable limits of this powerful technique.  The
second advantage is the fact that preparing the chromophore in a
surface location is much easier in a cluster than in a matrix. The two
disadvantages are (1) the usual transient character of every molecular
beam, which makes it necessary to use powerful laser sources to both
pump and probe the cluster beam, while the bulk matrix can be both
manipulated and probed with relatively low power, broadband light
sources, and (2) the fact that the temperature of a cluster is
``internally set'' by evaporative cooling and cannot easily be
changed.

Another problem connected with MI spectroscopy is the presence of
matrix-induced perturbations, which can be minimized by the use of
bulk solid or liquid helium as a medium. This, however, comes with a
heavy price: injection of the sample into helium without causing
aggregation or condensation onto the walls of the container is
extremely difficult because of low solubility, and has been achieved
almost exclusively for atoms~\cite{Tabbert97}.  It should be noted,
however, that high purity parahydrogen has been shown to form stable
matrices for molecular dopants with perturbations, as measured by
widths of rovibrational transitions, that are comparable to those
observed in helium~\cite{Tam99}.

The extension of CI spectroscopy to helium clusters, pioneered at
Princeton about a decade ago~\cite{Goyal92b}, solves the problem of
sample injection. It has grown tremendously in importance since the
group of J.~P.~Toennies in G\"ottingen has shown that a guest in a
host helium nanodroplet can yield a spectrum indicative of free
molecular rotation.
%
%The possibility of achieving rotational resolution couples the
%synthetic flexibility of MI with the structural sensitivity of
%rotational spectroscopy, propelling this technique to the forefront in
%the study of unstable species. Furthermore, it
Almost all the spectra measured to date are consistent with the
spectral structure predicted by the symmetry of the isolated molecule
(see Section~\ref{sec:molsym}).
% in other words the helium density (viewed as quantum probability
% density of a many body wavefunction, rather than a classical quantity,
% see Section~\ref{sec:}) must adapt to the symmetry of the interaction
% potential that the molecule creates for the helium.
In contrast, in a solid matrix (or even in a conventional liquid, if
an instantaneous configuration is considered), the total symmetry of
the system is the combined symmetry of the molecule and that of the
trapping site, usually lower than that of the molecule alone.
Furthermore, the matrix-induced spectral shifts in helium are also
much smaller than in any other medium, and often much smaller than
complexation-induced shifts (see Section~\ref{sec:shifts}). For the
purpose of comparison with \emph{ab initio} calculations, complexes
formed in He droplets can then be treated as isolated (gas phase), and
the observed shifts provide direct information to decide which isomers
are actually formed. There are two important implications: (1) the
structural information contained in the spectral symmetry is
undiluted, and (2) if a complex is formed in helium, the interactions
of its parts with the surrounding medium are too weak to significantly
affect its equilibrium structure (zero-point motions and, more
dramatically, kinetics, are sometimes affected).

Finally, the possibility of obtaining high resolution spectra in
helium droplets at 0.38~K and, using $^{3}$He, 0.15~K, allows for the
study of superfluidity in nanoscale systems, a frontier subject in
condensed matter physics that has the potential to give a substantial
contribution to our understanding of highly degenerate many-body
systems~\cite{Grebenev98,Lehmann98}.

\section{Experimental methods}
\label{sec:exptl}

\begin{figure}[!h]
\includegraphics{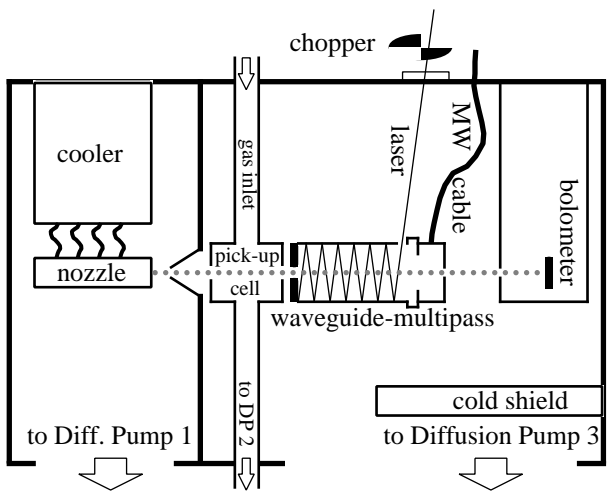}
  \caption{Schematic of a typical experimental apparatus. From
    Ref.~\onlinecite{Callegari00b}.}
  \label{fig:exptl}
\end{figure}

The history of liquid helium cluster/nanodroplet production and the
principal methods used for their study have been reviewed in the
preceding article by Jan Northby~\cite{Northby01_here}. Here we will
instead discuss the design criteria of a typical droplet source for
HENDI spectroscopy, with particular focus on those used in our
laboratory. The schematic of a typical experimental apparatus is shown
in Figure~\ref{fig:exptl}.

\subsection{Nanodroplet beam production}
\label{sec:production}

Due to the weak binding energy of small helium clusters, large
densities and low temperatures are called for in the expansion, both
resulting in a large gas flux.  Therefore, liquid helium droplet
sources require large pumps and/or very small nozzles.  The flux is
proportional to $P_0 d^2 T_0^{-1/2}$, where typical nozzle diameters
($d$) range from 5 to 20~$\mu$m, pressures ($P_{0}$) from 1 to
100~bar, and temperatures ($T_{0}$) from 10 to 30~K\@. Within the
ideal gas approximation, a 10~$\mu$m nozzle with $T_{0}=20$~K and
$P_{0}=15$~bar will discharge into the source chamber approximately
the same gas load as a room temperature 50~$\mu$m nozzle at 2~bar,
i.e., \mbox{$\approx 4$} standard cm$^{3}$/s \mbox{($= 0.16$~mmol/s)}.
This is the maximum gas load that can be applied to a typical
diffusion pump with an effective pumping speed of 4000~L/s (referred
to N$_{2}$, i.e., 10000~L/s for He) at the limit operating pressure
\mbox{($\approx$ \pt{3}{-4}~torr)}.  Typically, droplets with an atom
count $N$ of a few thousand are formed. The large mass of these
droplets means that the beam intensity is not much affected by
scattering from the background gas.%%%NEW

% This can be taken advantage of by using turbomolecular
% pumps, which have larger maximum operating pressures (up to a factor
% of 5) than diffusion pumps.
%
%This is only true if one takes into account in the design of the
%source that the thermal load on the cold source also increases
%accordingly, and may therefore become the limiting factor.  It can be
%shown that

%%%NEW
If a source is fed at the ``standard'' load mentioned above, it is
possible to operate a droplet source with a small two-stage
closed-cycle refrigerator (the warm stage typically operates at 80~K,
the cold stage at 12~K) delivering 1~W of cooling power at the cold
stage (0.12~W of power are needed to cool the gas from 80~K down to
20~K).  However, due to background radiation and background gas
thermal loads, the use of a 5~W refrigerator is recommended. This will
allow experiments to span a larger temperature and gas load range, and
therefore to cover a larger range of average cluster sizes.  For
sources which operate below 10~K, which are used for the production of
very large clusters ($\nbar > 10^{5}$), either refrigerators with
lower minimum operating temperatures or direct liquid He cooling are
required.  Closed-cycle refrigerators with a zero-load temperature of
3~K are commercially available~\cite{cryo}, of course at substantially
higher cost. Cost is also the limiting factor for direct liquid He
cooling, which is the cooling method that allows the maximum
flexibility in terms of droplet size range and total flux that can be
attained~\cite{Buchenau90}.

In a subcritical expansion, $^{4}$He droplets exhibit a log-normal
size distribution, with a full width at half maximum of
0.55--0.75~\nbar~\cite{Harms98a,Harms01}.  While the existence of
scaling laws for the determination of \nbar\ from expansion parameters
is generally recognized, different formulas have been proposed for
it~\cite{habe94}. The most popular one states that \nbar\ is a
function (not necessarily linear~\cite{Smith98}) of $P_{0} d^{1.5}
T_0^{-2.4}$.
% The difficulty of finding a unique and reliable scaling law reflects
% the fact that it is computationally prohibitive to exactly follow the
% beam evolution during the expansion, and several approximations have
% to be made (most notably, the use of ideal gas formulae and classical
% nucleation theory). A detailed account of these difficulties has been
% given by Knuth~\cite{Knuth97}, who also explains how an absolute
% calibration of scaling laws is complicated by the difficulty of
% experimentally determining cluster sizes (usually done by mass
% spectroscopy), because of fragmentation, and variable ionization and
% detection efficiency; while the use of scattering techniques can
% avoid some of these problems, it raises the issue of how efficiently
% the momentum of a foreign particle is transferred to a He droplet.
The estimates that we will give here for \nbar\ use a formula by Knuth
\etal~\cite{knut94}; this formula is in fair agreement with the
measurements of Ref.~\onlinecite{Harms98b}, although the latter seem
to scale differently with $T_{0}$.  The case of supercritical
expansion, which produces a bimodal distribution of much larger
droplets, has been analyzed in
Refs.~\onlinecite{Buchenau90,Harms97a,Knuth99}.  IR absorption has
been used as a calibrated source of He evaporation
%(see Section~\ref{sec:IR}),
to measure average cluster sizes as a function of source
conditions~\cite{Hartmann99}.

The exact shape (density distribution) of a He droplet has been
computed using both density functional~\cite{Harms98a,Harms01} and
Monte Carlo~\cite{Ceperley95} methods. A good approximation for ``back
of the envelope'' calculations is that of a sphere having the same
density as bulk liquid helium ($\rho_{0} = 0.0218$~\AA$^{-3}$, for
$^{4}$He~\cite{Abraham70}). According to this liquid drop model, a
droplet containing $N$ $^{4}$He atoms has a radius $R =
2.22N^{1/3}$~\AA\@. A more realistic description uses the analytic
function~\cite{Harms98a,Jortner92}:
\begin{equation*}
  \label{eq:density}
  \rho(r)=\frac{\rho_{0}}{2}\left[1-\tanh\left(2\frac{r-R}{g}\right)\right],
\end{equation*}
which accounts for the diffuseness of the surface via a thickness
parameter $g$; the 10--90\% thickness \mbox{($\approx 1.1 g$)} is
estimated by various methods to be \mbox{$\approx 6$~\AA} for
$^{4}$He~\cite{Harms98a} and \mbox{$\approx 8$~\AA} for
$^{3}$He~\cite{Harms01}.

The equilibrium temperature of a droplet is dictated by evaporative
cooling~\cite{Gspann82,Brink90}, and can be extracted from the fit of
the ro-vibrational spectrum of a probe molecule (Sec.~\ref{sec:IR}).
The measured values, 0.38~K for $^{4}$He (Ref.~\onlinecite{Harms97b},
and most entries in Tab.~\ref{tab:molecules}) and 0.15~K for
$^{3}$He~\cite{Harms97b}, are in good agreement with theoretical
predictions~\cite{Gspann82,Brink90}.

% For the production of H$_{2}$ clusters,
%keeping the nozzle diameter and the pressure constant, the temperature
%corresponding to 20~K needed to produce the ``standard' droplet beam
%is approximately 72~K\footnote{This estimate is made assuming that the
%  same reduced $T^{*}$ needs to be reached, where $T^{*} =
%  \frac{kT}{\epsilon}$ and $\epsilon$ is the well depth of the
%  potential (10~K for He, 36~K for H$_{2}$)}, which is not far from
%what can be easily reached by liquid N$_{2}$ cooling.

About 1~cm downstream from the nozzle, the droplet beam is
collimated by a skimmer.
%
%The purpose of the skimmer is to admit into
%the detection chamber only the useful center portion of the beam,
%while keeping most of the gas load in the source chamber. This results
%in low background pressure, which  minimizes attenuation of the
%droplet beam due to scattering.
%
As shock waves may play a role in attenuating the beam, it is
important that the shape of the skimmer is optimized, and that its
edges are as sharp as possible. Typically, commercial electroformed
conical skimmers, about 500~$\mu$m in diameter, are
used~\cite{beamdynamics}.

To give an idea of what intensity can be expected for a droplet beam,
consider a detector located 500~mm away from the source, with a
1~mm$^2$ detection area, and an average cluster size of $\nbar \approx
4000$.  Assuming that 10\% of the flux is in the droplets and that the
droplet flux is spread uniformly over 1/10 of a steradian in the
forward direction (a conservative estimate), one gets a flux of on the
order of $10^{11}$ droplets per second at the
detector~\cite{Miller88}.  Notorious difficulties in calibrating the
absolute values of molecular beam detectors (be it bolometers or mass
spectrometers) make this number hard to calculate precisely.  For
larger clusters, a first approximation of the observable mass flux can
be obtained by scaling the number of clusters inversely with the
number of atoms per cluster.

Whereas pure and doped $^{4}$He droplets have been extensively studied
far fewer experiments with pure $^{3}$He or mixed $^3$He/$^4$He
droplets have been done~\cite{Stephens83,Gspann95,Farnik98,Harms99b,%
  Harms01,Harms97b,Harms99a,Grebenev98} chiefly due to the high cost
of the isotope~\cite{isotec}.  Budget dictates that $^{3}$He be
continuously recycled, requiring specialized equipment: sealed pumps
and a purification system. The gas is collected from the exhaust of
the source chamber pumping system, cleaned on liquid-nitrogen-cooled
zeolite traps, pressurized by a membrane compressor to 20--80 bar and
fed back to the nozzle. The recycling system used in
Refs.~\onlinecite{Farnik98,Harms99b,Harms01,Harms97b,Harms99a} needed about
20 bar$\cdot$L gas to be filled.  The loss rate under normal operation was
$< 2$ bar$\cdot$L per week~\cite{Harms99a}.  Thus although a considerable
initial investment is required the operating costs are comparable to
the other running costs of these experiments.

There is an important difference between $^{4}$He$_{N}$ and
$^{3}$He$_{N}$ clusters: whereas the former are bound for any
$N$\cite{Schollkopf94}, calculations predict that small $^{3}$He
droplets of less than about 30 atoms are
unstable~\cite{Guardiola00,Barranco97} because of the larger zero
point energy associated with the lighter isotopic mass.
Experimentally, a strikingly different behavior of average droplet
size as a function of nozzle temperature is observed for the two
isotopes. Whereas the size of $^{4}$He droplets increases gradually
with decreasing temperature\cite{Harms98a}, the formation of $^{3}$He
droplets sets in sharply, and at a considerably lower temperature
($T_{0} = 11.5$~K for $P_{0}=20$~bar and $d=5 \mu$m, giving droplets
of about 4500 atoms \cite{Harms01,Harms99a}).  In the expansion of an
isotopic mixture, formation of $^{4}$He-enriched mixed droplets is
observed~\cite{Harms97b,Harms99a}.  The $^{4}$He concentration in the
droplets is several times higher than that of the expanding mixture
($<4\%$) and depends on the expansion conditions~\cite{Harms99a}.
Because $^4$He will selectively solvate most dopants, the production
of pure $^{3}$He droplets requires gas of high isotopic purity
($^{4}$He $<10^{-6}$), which is commercially available~\cite{isotec}.

\subsection{Nanodroplet beam doping}
\label{sec:pickup}

After collimation, the droplets are seeded with the molecule of
interest by passing them through a small gas pickup cell where the
pressure is tailored to obtain the desired mean number of dopants per
droplet.  The internal and kinetic energy of the dopant, as well as
the energy of solvation, are dissipated into the droplet and result in
evaporation of He atoms (200--250 atoms, each taking off 5.5--7~K of
energy~\cite{chin95}, are estimated for the pickup of a small molecule
such as HCN), leading to a microcanonical system cooled to $0.38$~K
(we assume pure $^4$He droplets here). On average, the impact
parameter of the molecule is significant (2/3 of the droplet radius),
resulting in a mean angular momentum of \mbox{$\approx 45 N^{1/3}
  \hbar$} per collision with HCN\@. Likewise, He evaporating at 0.38~K
will carry away 10--15$\hbar$ of angular momentum per atom. At
present, it is not clear how much angular momentum remains in the
cluster after evaporative cooling, nor in which form it is stored if
it exceeds the thermal values.

A quick estimate of the optimum pickup conditions can be made by
assuming unity sticking coefficient and a droplet cross section of
15.5$N^{2/3}$~\AA$^{2}$ (see Section~\ref{sec:production}); further,
the droplet's size reduction upon pickup and the nonzero velocity of
the gas molecules are neglected (if the latter is accounted for, the
collision rate will increase by a factor of
\mbox{$\approx\sqrt{1+(v\si{m}/v\si{d})^2}$}, where $v\si{d}\approx
400$~m/s is the droplet velocity and $v\si{m}$ is the rms velocity of
the molecules in the pickup cell).
%~\cite{goya92,goya93,Gough85}.
With these approximations (further corrections are discussed in
Ref.~\onlinecite{Lewerenz95}), the pickup process results in a Poisson
distribution of the number of dopant molecules per droplet, and a
column density corresponding to \mbox{$\approx 2.5
  N^{-2/3}$~Pa$\cdot$cm} is required to maximize the probability of
single pickup.  Under typical conditions, the vapor pressure needed
for efficient doping of helium droplets is on the order of
10$^{-2}$~Pa, some four orders of magnitude lower than what is
typically used in supersonic jet coexpansions.  This implies that many
species can be doped into helium that are not practical to study by
jet spectroscopy because of the high temperatures involved, which are
either difficult to reach or cause thermal decomposition of the
species.  The large cross sections for the droplets imply that the
vacuum requirements in the instruments are more stringent than for
beam experiments on monomers or small clusters.

Molecules captured by the same droplet are expected to find each other
and form van der Waals complexes, but once past the pickup region,
each droplet becomes an isolated system and no further aggregation
occurs.  At the temperature of the droplet, 0.38~K, the ro-vibrational
spectrum of a typical molecule spans \mbox{$\sim 1$~\wn}, which is
generally smaller than the shifts induced by complexation.  Hence,
spectra of different small oligomers will not overlap, and pickup of
more than one molecule per droplet will only decrease the intensity of
the single-molecule spectrum, without affecting its shape. For this
reason, at least when one is concerned with monomers and dimers only,
high purity of the dopant gas is desirable but not essential. An
extreme case is the spectrum of H$^{12}$C$^{13}$CH, which has been
observed~\cite{Nauta_preprint3} at its naturally occurring abundance
(2.2\%) interleaved to that of H$^{12}$C$^{12}$CH\@.  The situation is
different for higher oligomers, because complexation shifts tend
towards an asymptotic limit~\cite{Nauta99a,Nauta99d}, and the spectra
tend to merge. Nevertheless, in a favorable case complexes of OCS with
up to 17 hydrogen molecules have been made and spectroscopically
resolved~\cite{Grebenev00c}, though this was done in mixed
$^3$He/$^4$He droplets which cool to 0.15~K\@. Pendular spectroscopy
(see Section~\ref{sec:stark} and Ref.~\onlinecite{Nauta99a}) allows
even the modest width of the rotational contours to be collapsed,
which is particularly favorable in separating closely spaced oligomer
or isomer bands.
%
%It should be noted that the latter
%experiment took advantage of the colder temperature of mixed
%$^{3}$He/$^{4}$He droplets. The standard approach to resolving
%closely-spaced bands is, when applicable, pendular spectroscopy,
%presented in Section~\ref{sec:stark}.

\subsection{Radiation sources}
\label{sec:lasers}

Since spectroscopy of doped helium droplets is still a relatively new
field, it is often the availability of a suitable radiation source
that dictates the choice of the system to be studied. In the early
days~\cite{Goyal92b,Goyal93b}, SF$_{6}$ (\vi{3} mode) was chosen as a
dopant molecule because of its large absorption cross section in a
spectral region where watts of radiation (from line-tunable CO$_{2}$
lasers) were available.

To achieve rotational resolution, this transition was later studied
with a continuously tunable semiconductor (lead salt)
laser~\cite{Hartmann95}. The same type of laser was also employed for
the beautiful studies of OCS in mixed $^{3}$He/$^{4}$He carried out in
G\"ottingen~\cite{Grebenev98,Grebenev01,Grebenev01b,Grebenev00c}.

Pulsed lasers have also been used in early experiments on HF-,
H$_{2}$O- or NH$_{3}$-doped helium
droplets~\cite{Blume96,Frochtenicht96,Huisken99}. They offer the
advantage of high power and wide-range tunability; those come at the
price of usually broadband emission, so that it becomes difficult to
establish whether the width of the observed spectral lines is
determined by the surrounding helium environment or rather by the
laser itself (via instrumental linewidth or power broadening).

High power line-tunable gas lasers, cw or pulsed, are still being used
up to the present time~\cite{Behrens97,Behrens98,Behrens99,Kunze00}.
Because of the sparse set of available lasing frequencies, they are
most suited to the investigation of small molecules, with small
moments of inertia and therefore well spaced lines. Thanks to their
large output power, they have been successfully used to perform
saturation measurements~\cite{Kunze00}.

Most of the experimental spectra to date have been acquired with color
center lasers, in the 3~$\mu$m
region~\cite{Nauta99a,Nauta99b,Nauta99d,Nauta00a,Nauta00b,Nauta00c,%
  Nauta01a,Callegari00b}, and, to a lesser extent, in the 1.5~$\mu$m
region~\cite{Callegari00c}.  These lasers offer broader tunability
range, comparatively large output power (30 and 250~mW, respectively)
and high resolution, as they can be stabilized to better than 1~MHz
without too much difficulty. Since the 1.5~$\mu$m region is used in
telecommunications, commercial optical components (e.g., fibers,
connectors, and high finesse cavities) are available for this type of
measurements.

At these power levels it is possible to bring near saturation a 10~MHz
homogeneously broad, strong fundamental transition near 3~$\mu$m. If
the lines are broader or the transition weaker (as in the case of
overtones at 1.5~$\mu$m), laser field enhancement devices can be
usefully employed (see Section~\ref{sec:excitation}). Looking at the
future, since HENDI spectral lines are rarely narrower than 0.01~\wn\
(see Section~\ref{sec:relax}) and since pulsed lasers with similar
bandwidth are available, it would probably be useful to explore the
possibility of developing pulsed helium nanodroplet sources.  Pulsed
devices that produce small He clusters are already available at
present~\cite{even00}.

For cw beam sources, narrow bandwidth high repetition rate pulsed
lasers have recently become available at power levels that allow
frequency multiplication and mixing. As the use of regenerative
amplifiers has allowed repetition rates up to 10~kHz, the problem of
the low duty cycle characteristic of traditional pulsed laser sources
can be overcome. For strictly cw laser sources, recent developments
that will likely find HENDI applications are cw OPOs~\cite{Powers98}
and quantum cascade lasers~\cite{Capasso99}.  Quantum cascade lasers
offer the advantage of being available to wavelengths as high as
\mbox{$\approx 24$~$\mu$m}~\cite{Colombelli01}, which is going to be
particularly useful for the study of large molecules where, at higher
frequencies, Intramolecular Vibrational Redistribution (IVR) is an
unavoidable source of line broadening and of loss of information.

\subsection{Excitation schemes}
\label{sec:excitation}
\subsubsection{Infrared}
\label{sec:IR}

\begin{figure}[!h]
\includegraphics{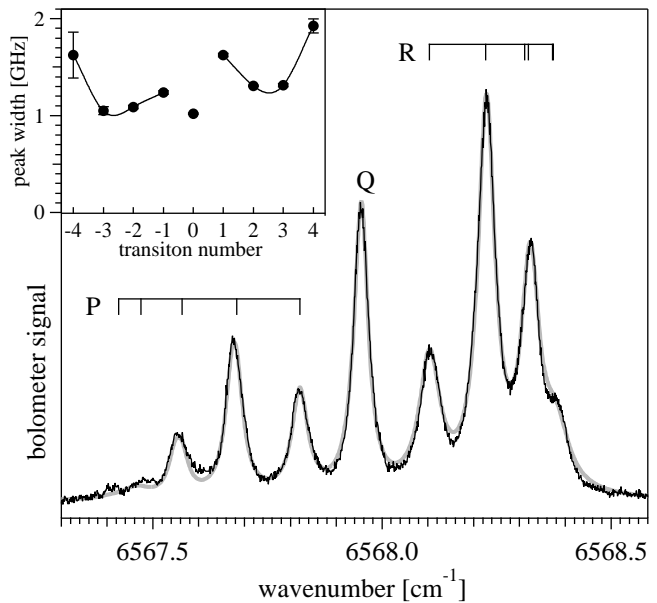}
  \caption{Spectrum of the 2\vi{1} transition of propyne, and fit to a
    symmetric top. The inset shows the linewidth for each transition.
    Note that at the temperature of the droplet (0.38~K) the Q-branch
    would be absent if nuclear spin relaxation had taken place. From
    Ref.~\onlinecite{Callegari00c}.}
  \label{fig:propyne}
\end{figure}

Infrared (IR) spectroscopy in He droplets derives its uniqueness from
the fact that the medium still allows rotational resolution (a
representative spectrum is shown in Figure~\ref{fig:propyne}), even
for relatively large molecules and complexes.  This is to be compared
to ``classical'' solvents, for which rotational resolution is, at
best, only possible for molecules of the smallest moments of inertia.
Rotational resolution allows the measurement of the cluster
temperature, a very sought after quantity as the following statement
(made in 1994) proves:
\begin{quote}
But so far no ``thermometer'' to measure a cluster's temperature is
available. Worse, there does not even exist a suggestion how to
construct one.\\
{[Ref.~\onlinecite{Haberland94}, Chapter 1.5: \textit{Experiments
  not possible today}]}
\end{quote}
Ironically, in the same book, the first results on infrared
 spectroscopy in He clusters were summarized, and the following
 prediction was made:
\begin{quote}
In future the technique may be combined with continuously tunable
narrow band lasers of higher power as these become available.\\
{[\textit{ibid.}, Chapter 3.7: \textit{Infrared spectroscopy}]}
\end{quote}

Continuously tunable, narrow-band lasers were indeed the key that, one
year after, allowed for rotational resolution to be achieved and thus
temperature to be measured~\cite{Frochtenicht94,Hartmann95,Harms97b}.
The technical solution that was used to overcome the power requirement
consists of a different arrangement of how the laser beam interacts
with the beam of doped droplets. In contrast to the opaque bolometer
detectors of the early experiments, the use of a ``transparent'' mass
spectrometer detector allowed for a setup where the laser and cluster
beam counterpropagated, resulting in interaction region of tens of
centimeters. With this arrangement, low power diode lasers can be
profitably employed~\cite{Hartmann95}. Additional benefits of using
this setup for pulsed lasers is that their duty cycle (defined as the
fraction of the beam that is in the laser field) is increased from
$10^{-5}$ to $10^{-3}$ and the detector output can be time gated to
match the depletion transient.

When bolometers are used as detectors, multipass cells~\cite{Nauta99a}
and power build-up cavities~\cite{Callegari00c} have been successfully
employed. While a well aligned multipass cell can give a 30- to
50-fold increase in signal relative to the single-crossing case, a
resonant cavity can further enhance the radiation intensity seen by
the droplets by one order of magnitude.  Naturally, this comes at a
price, since, as the laser is scanned, the cavity has to be kept in
resonance by suitable feedback electronics.  Moreover, as currently
implemented, the interior of the cavity is a volume bound by metallic
surfaces; as such it is effectively shielded from electromagnetic
fields. Hence, double resonance and Stark spectroscopy experiments are
not possible.  Nevertheless, the large amount of available power
(currently only matched by line-tunable gas lasers~\cite{Kunze00})
makes this approach very desirable to reach saturation conditions, and
to study forbidden transitions~\cite{Callegari00c}.
%
%Particularly if a comparison between the corresponding spectra is
%possible, a large amount of information can be extracted from the
%shape and saturation behavior of MW and IR lines. A detailed account
%is given in Section~\ref{sec:relax}. In brief, it is found that vibrational
%relaxation almost never contributes significantly to the linewidth and
%that exceptions are associated to intramolecular effects such as Fermi
%resonances and predissociation. Whether or not rotational relaxation
%contributes mostly depends on the coupling to the modes of the He
%bath: molecules with rotational transitions at energies near or above
%bulk modes (rotons) do exhibit lifetime broadened lines ($>$ xx \wn);
%all others, which have to relax into surface modes (ripplons) exhibit
%inhomogeneously broadened lines, although the homogeneous linewidth
%can still be measured through saturation behavior, and further
%information can be extracted via double resonance experiments.

HENDI spectra have been measured by beam depletion using both
bolometers and mass spectrometers to detect the on axis beam flux.
Even with helium evaporation occurring at 0.38~K, the mean lab frame
scattering angle of the evaporated helium is still $\approx 10$
degrees, and so the evaporated helium largely miss the detector.  The
relative sensitivity of the two detection methods is hard to determine
unambiguously from the literature spectra, as they involve widely
different source powers, transition moments, and detection geometries.
Our attempt to correct for these factors suggests that detection of
depletion with a bolometer is about a factor of ten more sensitive
than with a mass spectrometer. This assumes use of a continuous wave
laser and of a multipass cell. The latter partially offsets the
shorter excitation length (relative to the more efficient counter
propagation scheme possible with the mass spectrometer) so that
ultimately only modest reduction results.  The mass spectrometer will
likely give superior sensitivity in the case when one is using a low
repetition rate pulsed laser for excitation, due to the possibility of
using gated detection.

\subsubsection{Microwaves}
\label{sec:MW}

Investigation of purely rotational transitions in doped helium
clusters shifts the frequency spectrum to the microwave (MW) region.
Radiation is confined inside a section of waveguide (part of which is
collinear with the cluster beam) properly terminated at the ends.
Confinement is necessary not only to concentrate the available
microwave power, but also to minimize its absorption or rectification
by the bolometric detector, which can easily obscure the true signal.

%Pure microwave spectroscopy in He droplets sets itself apart from all
%other spectroscopic techniques in that it has no gas phase
%counterpart, and is rather the microscopic equivalent of microwave
%cooking.
%
%One must remember that in a supersonic beam, it is unfeasible
%to measure directly the attenuation of exciting radiation, as it is
%unfeasible to detect the reemitted photons, which is why the absorbed
%energy is monitored instead, either directly or via the induced
%dissociation of the absorbing unit; electric-resonance, i.e.,
%photon-induced change in focusing by field gradients, has also been
%used~\cite[and refeences therein]{klem95}.
Pure rotational spectroscopy in He droplets differs from vibrational
and electronic spectroscopies in that it is inherently multiphoton.
One can see from Table~\ref{tab:molecules} that in a He droplet the
energies associated with a pure rotational transition (0.5--70 GHz)
are insufficient to evaporate even a single He atom.  Microwave
spectroscopy in a He droplet relies on the occurrence of rotational
relaxation in an isolated droplet, which allows a dopant molecule to
repeatedly absorb photons and release their energy into the droplet.
The fact that the observed signals are of the same magnitude as for
infrared transitions has been used to estimate the number of
relaxation cycles that each molecule must undergo in the
\mbox{$\approx 100$~$\mu$s} it spends in the MW field, and thus to
place the rotational relaxation time (see Section~\ref{sec:relax}) on
the sub-microsecond time scale for HCCCN~\cite{Reinhard99}.

The large amount of available microwave power (tens of watts), along
with the large transition dipoles in this spectral region has made
saturation measurements
possible~\cite{Reinhard99,Conjusteau00,Callegari00b}, which have been
used to establish and quantify the dominance of inhomogeneous
broadening in HENDI spectroscopy.

\subsubsection{Double resonance}
\label{sec:DR}

The term double resonance (DR) refers to the coupling of two
transitions by a common level, which allows driving one transition
(the ``pump'') to change the strength of the other (the ``probe'').
In the gas phase, DR is routinely used (1) to simplify the assignment
of complex spectra, (2) to reach states inaccessible by one-photon
selection rules, and (3) to measure homogeneous lines within an
inhomogeneous profile.  To date, with one exception, only type (3)
experiments have been done in helium droplets, and only in the
simplest ``steady state'' versions, i.e., without explicit time
resolution of the time delay between pump and probe fields.  The
exception is the recent hole burning study of the electronic spectrum
of tetracene~\cite{Hartmann01} which shows a helium induced splitting
of the electronic band origin.  The complexity of the dynamics in
helium droplets has unfortunately made the interpretation of these
experiments much more difficult than that of their gas phase
analogues.

% In a droplet, the homogeneous
% broadening of the spectrum can in principle contain contributions from
% the rates of spectral diffusion, rotational population relaxation, and
% pure ``dephasing''. The latter arises from elastic scattering between
% the impurity and the thermal excitations of the droplet, predominantly
% surface ripplons.

% The first double resonance study in helium was the MW-MW set of
% experiments performed on the R(3) and R(4) transitions of HCCCN.
% Saturation measurements on these rotational lines had demonstrated
% that their strength was proportional to the square root of the applied
% power over a wide range, the classic signature of saturation of a line
% dominated by inhomogeneous broadening. It was expected that the DR
% would directly reveal the homogeneous component of the line via
% hole-burning with the the pump field, i.e., a Bennet hole and hill in
% the absorption of the probe~\cite{ref}.  Based upon analysis of the
% single color saturation study, in the limit of \emph{static}
% inhomogeneous broadening a DR width of no more than 150~MHz was
% expected.  It was instead found that the linewidths of the transitions
% affected by the pump beam were comparable to those measured in single
% resonance (1~GHz). This observation was ascribed to the spectral
% diffusion of the dopant over the quantum states responsible for the
% inhomogeneous broadening (most likely different translational states,
% see Section~\ref{sec:translation}) occurring on a similar time scale as
% rotational population relaxation (see Section~\ref{sec:relax}).

The first double resonance study in helium were the MW-MW experiments
performed on the R(3) and R(4) transitions of HCCCN~\cite{Reinhard99}.
Saturation measurements on these rotational lines had shown the
typical pattern of a line to be dominated by inhomogeneous broadening.
It was expected that DR spectra would directly reveal the homogeneous
component of the line, as in a classical hole-burning experiment
(Ref.~\onlinecite{Demtroeder96}, page 438).  The observation of lines
much broader than expected was interpreted as a sign of spectral
diffusion (see Section~\ref{sec:relax2}).

In an attempt to gain further insight, MW-IR experiments on
HCCCN~\cite{Callegari00b} have been performed.  Independent of this
work, MW-IR double resonance experiments where performed on
OCS~\cite{Grebenev00b}.  Because rotational relaxation times in He
clusters lie in the nanosecond range (during which a cluster only
travels a few $\micro$m), double resonance experiments require spatial
overlap of the two radiation fields. While in MW-MW experiments this
is readily achieved, in MW-IR experiments spatial overlap has been
achieved by either propagating the laser along the axis of the
waveguide~\cite{Grebenev00b}, or by using the walls of the waveguide
as a substrate for thin mirrors in a multipass
arrangement~\cite{Callegari00b} (see Figure~\ref{fig:exptl}).  The
long wavelength of microwaves, compared to the transverse dimensions
of the cluster beam, makes it possible to have holes drilled into the
waveguide as ports for the cluster and laser beam, without introducing
significant microwave losses.

While a MW-IR scheme introduces the additional complication of
vibrational excitation (which fans out the ``probe'' transition
frequencies associated to a given ``pump'' frequency), it has the
advantage that the probe laser can cover the entire manifold of
rotational levels, thus probing rotational states that are not
directly connected via the pump field. With some differences, probably
due to the use of different molecules, both experiments find that the
population of \emph{all} levels is to some extent affected by the pump
beam, which indicates fast rotational diffusion.

\subsubsection{Stark spectroscopy}
\label{sec:stark}

Stark spectroscopy has been shown to be a potentially valuable tool to
learn more about the interactions of the dopant molecule with the
helium environment and possible distortions of the molecule
(particularly of complexes) by gentle helium
solvation~\cite{Nauta99a,Nauta99b}.  In a simple linear dielectric
approach, one would expect the molecular dipole moments to be reduced
in helium by its dielectric constant, 1.055~\cite{crc70}.  However,
this is only valid for a dipole in a continuous medium. If the dipole
is located in the center of a spherical cavity, the induced dipoles in
the solvent cancel out and give no net change in dipole moment.  As
the solvation around molecules is in general highly anisotropic, the
net induced moment provides a potentially interesting measure of the
anisotropy. Recently such calculations have been performed for Mg (3P)
solvated in He, with the anisotropic density obtained by a Density
Functional method, and a 6.2\% reduction of the optical transition
dipole of Mg (3P $\rightarrow$ 3S) transition has been
calculated~\cite{Reho00a}.  Such a reduction, along with the
solvent-induced red shift, has been found to quantitatively account
for the experimentally measured increased lifetime of the transition.
The solvent-induced stabilization of dipolar structures was listed as
one of the possible causes of the decreased tunneling splitting
observed for the HF dimer in He droplets (see Section~\ref{sec:nh3}).

Nauta and Miller~\cite{Nauta99b} have studied the Stark pattern of the
R(0) line of HCN in the IR\@.  They find the splitting and intensity
pattern expected from the isolated molecule, but with the observation
that what appears to be the a splitting of the $|M| = 0,1$ components
persists even at zero field, as a poorly resolved doublet structure of
the R(0) line~(see Figure~\ref{fig:hcn}).  A careful study of the
dipole moment in this and other systems would likely be quite valuable
for testing theoretical predictions of the helium solvation structure.

Another relevant manifestation of Stark effects is the mixing of
rotational states (Ref.~\onlinecite{Townes75}, Section~10.6). In He
droplets, it has been exploited to induce a (normally forbidden)
Q-branch in those linear molecules, such as HCN and HF, whose
zero-field rotational constants are large enough that their spectrum
consists of the R(0) line only. By this method, the band origin and
the rotational constant of the molecule can be directly
measured~\cite{Nauta99b}.  In this application the field is kept as
weak as possible, compatibly with the need of inducing a measurable
signal, and indeed an accurate value for the band center is obtained
by measuring the position of the Q-branch at increasingly weak fields
and extrapolating to zero field.

The electric field can be used to ``tune'' the frequency of a given
transition. This has been exploited to quench tunneling in
(HF)$_{2}$~\cite{Nauta00b}, and to perform time-resolved relaxation
measurements in the ms range (by mapping different excitation times
onto an electric field gradient)~\cite{Nauta_preprint2}.

Once the information contained in the ro-vibrational Hamiltonian has
been extracted, rotational resolution actually becomes a nuisance:
first, the band strength is fractionated, resulting in smaller
signals; second, a typical spectrum spans a range comparable to
$k\si{B}T$ (0.38~K or 0.26~\wn), which results in a corresponding loss
of resolution when trying to identify closely spaced bands (a common
occurrence when many different van der Waals complexes of the same
molecules are present).  For gas phase polar molecules, an elegant
solution is Pendular State spectroscopy~\cite{Loesch90,Friedrich91},
which is the application of a static electric field sufficiently
strong to convert the rotation of a molecule into oscillations of the
dipole moment around the external field.  In the strong static field
limit (and with the permanent and transition moments parallel to each
other), the entire rotational contour of the spectrum collapses into a
single line at the wavenumber of the purely vibrational excitation,
with an integrated intensity 3 times that of the entire band at zero
field (the traditional spectroscopic ``conservation of intensity''
holds, it is just the orientation of the molecules is now fixed, and
absorption, no longer averaged, will be maximum when the transition
dipole and the laser polarization are parallel).  It is indeed possible
to use the signal strength as a function of relative polarization of
the optical and Stark fields to determine the angle between the static
and transition dipoles.

\begin{figure}[!h]
\includegraphics[angle=90]{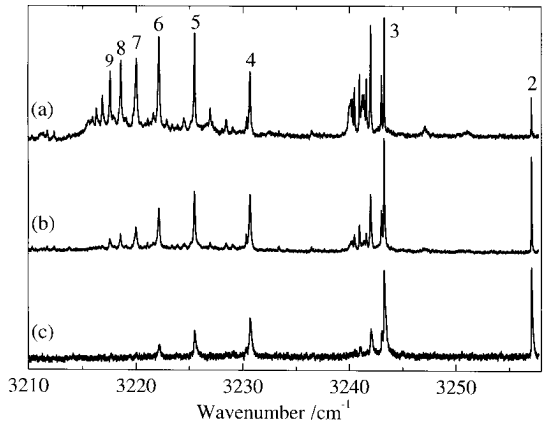}
  \caption{Stark spectrum of linear chains of cyanoacetylene (labeled
    by the number of monomer units), showing the high resolution
    achievable with pendular state spectroscopy. The series is taken
    at progressively smaller droplet sizes, showing that the size of
    the droplet limits the length of the chains that can be formed.
    From Ref.~\onlinecite{Nauta99d}.}
  \label{fig:stark}
\end{figure}

This technique has been ported to He droplets by Nauta and
Miller~\cite{Nauta99a,Nauta_Campargue}. It is particularly attractive
in this environment because both the low temperature and the increased
moments of inertia make it easier to reach the required ``high field''
limit.  By this technique, the spectra of linear chains of up to 8 HCN
units~\cite{Nauta99a}, and up to 12 HCCCN units~\cite{Nauta99d} have
been resolved (see Section~\ref{sec:noneq} and
Figure~\ref{fig:stark}).

Suppression of rotation can also be achieved by replacing $^{4}$He
with $^{3}$He~\cite{Grebenev98,Grebenev01}.  The technique has been
used to collapse the absorption bands of OCS--(H$_{2}$)$_{n}$
complexes so that $n$ could be assigned via the measured band
shift~\cite{Grebenev00c,Grebenev01b}.  This method is applicable even
to nonpolar cases, but lines will typically not be as narrow as for
$^4$He spectra.

\section{Current understanding and open questions}
\label{sec:openquestions}

\subsection{Molecular symmetries in the nanodroplet environment}
\label{sec:molsym}

%\begin{longtable}{lcllc}
\begin{table}[!hb]
  \caption{Rotational constants derived from rotationally resolved IR
    or MW spectra. Blank entries in the ``Mode'' column indicate purely
    rotational spectra. The modes in parenthesis indicate that more
    than one band has been seen but the measured rotational constants
    do not differ enough to grant separate entries.}
\begin{tabular}{lcllc}\hline\hline
Molecule & Mode & $B_{\mathrm{gas}}$ & $B_{\mathrm{He}}$ & Ref. \\
 &  & \wn & \wn &  \\ \hline
DCN & -- & 1.208 & 0.999 & \cite{Conjusteau00} \\
HCN & -- & 1.478 & 1.204 & \cite{Conjusteau00} \\
HCN & \vi{1} & 1.478 & 1.175 & \cite{Nauta99b} \\
(HCN)$_{2}$ & \vi{1} & 0.0582 & 0.0193 & \cite{Nauta99c}
\\
(HCN)$_{3}$ & \vi{1} & 0.0156 & 0.0073 & \cite{Nauta_HCN3}
\\
HCCCN & \vi{1} & 0.1517 & 0.0525 & \cite{Callegari00b}
\\
HCCCN & 2\vi{1} & '' & 0.0518 & \cite{Callegari00c}
\\
(HCCCN)$_{2}$ $(A)$& \vi{1} (\vi{2}) & -- & 0.0168 & \cite{Nauta99d}
\\
(HCCCN)$_{2}$ $(B)$& '' &  0.0113 & 0.0050 & \cite{Nauta99d}
\\
HCCCCH & \vi{3} & 0.1464 & 0.047 & \cite{Nauta_HCN3}\\
H$^{12}$C$^{12}$CH & \vi{3} & 1.1766 & 1.0422 &
\cite{Nauta_preprint3}\\
H$^{13}$C$^{13}$CH & \vi{3} & 1.1195 & 0.987 & \cite{Nauta_preprint3}\\
CH$_{3}$CCH & \vi{1} & 0.2851 & 0.0741 & \cite{Nauta_Campargue}
\\
CH$_{3}$CCH & 2\vi{1} & '' & 0.0717 & \cite{Callegari00c}
\\
CF$_{3}$CCH & 2\vi{1} $(A)$& 0.1908 & 0.1004 & \cite{Callegari00c}
\\
CF$_{3}$CCH & 2\vi{1} $(B)$& 0.0960 & 0.0355 & \cite{Callegari00c}
\\
HCN--N$_{2}$ & \vi{1} & 0.0525 & 0.019 & \cite{Nauta_HCN3}
\\
HF & \vi{1} & 19.787 & 19.47 & \cite{Nauta00c}
\\
(HF)$_{2}$ & \vi{1} (\vi{2}) & 0.2167 & 0.0986 & \cite{Nauta00b}
\\
CO--HF & \vi{1} & 0.524 & 0.1022 & \cite{Nauta_HCN3}
\\
N$_{2}$--HF & \vi{1} & 0.1066 & 0.045 & \cite{Nauta_HCN3}
\\
OC$^{32}$S & \vi{3} & 0.2029 & 0.0732 & \cite{Grebenev00a}\\
OC$^{34}$S & \vi{3} & 0.1979 & 0.0706 & \cite{Grebenev00a}\\
\pH--OC$^{34}$S $(A)$ & \vi{3} &      & 0.0847 & \cite{Grebenev01} \\
\pH--OC$^{34}$S $(B)$ & \vi{3} &      & 0.0544 & \cite{Grebenev01} \\
\pH--OC$^{34}$S $(C)$ & \vi{3} &      & 0.0422 & \cite{Grebenev01} \\
SF$_{6}$ & \vi{3} & 0.0911 & 0.034 & \cite{Hartmann95}\\
 HCOOH $(A)$ & \vi{2} (\vi{1}) & 2.5758 & 1.3811 & \cite{Madeja_preprint}\\
 HCOOH $(B)$ & '' & 0.4020 & 0.2970 & \cite{Madeja_preprint}\\
 HCOOH $(C)$ & '' & 0.3470 & 0.1996 & \cite{Madeja_preprint}\\
 DCOOH $(A)$ & \vi{1} & -- & 1.16 & \cite{Madeja_preprint}\\
 DCOOH $(B)$ & '' & -- & 0.24 & \cite{Madeja_preprint}\\
 DCOOH $(C)$ & '' & -- & 0.21 & \cite{Madeja_preprint}\\
CO$_{2}$ & \vi{3} & 0.39 & 0.154 & \cite{Nauta_preprint4}\\
NH$_{3}$ & \vi{2} & 9.96 & 7.5 & \cite{Behrens98}\\
\oH--HF & \vi{1} & 0.8325 & 0.182 &
\cite{Nauta_HCN3} \\
\pD--HF & \vi{1} & 0.492 & 0.164 &
\cite{Nauta_HCN3} \\
\oD--HF & \vi{1} & 0.487 & 0.159 &
\cite{Nauta_HCN3} \\
CH$_{4}$ & \vi{3} & 5.2406 & 5.013 & \cite{Nauta_HCN3}\\
N$_{2}$O & \vi{1}+\vi{3} & 0.419 & 0.0717 & \cite{Nauta_preprint4}\\
(CH$_{3}$)$_{3}$SiCCH $(A)$& 2\vi{1} & 0.1056 & 0.0275 &
\cite{Callegari00c}\\
(CH$_{3}$)$_{3}$SiCCH $(B)$& 2\vi{1} & 0.0654 & 0.0144 &
\cite{Callegari00c}\\
Ar--HF & \vi{1} & 0.102 & 0.04 & \cite{Nauta_HCN3} \\
Cyclopropane $(A+B)$&  & 0.6702 & 0.1631 &
\cite{Nauta_HCN3} \\
Cyclopropane $(C)$ &  & 0.4188 & 0.2416 & \cite{Nauta_HCN3} \\
\oH--HCN & \vi{1} & 0.4303 & 0.0897 &
\cite{Nauta_HCN3} \\
\pD--HCN & \vi{1} & 0.2663 & 0.0785 &
\cite{Nauta_HCN3} \\
Mg--HCN & \vi{1} & -- & 0.0285 &
\cite{Nauta01a} \\
Mg$_{3}$--HCN $(A)$ & \vi{1} & -- & 0.035 &
\cite{Nauta01a} \\
Mg$_{3}$--HCN $(B)$ & \vi{1} & -- & 0.0167 &
\cite{Nauta01a} \\
\hline\hline
\end{tabular}
\label{tab:molecules}
\end{table}
%\end{longtable}

The most remarkable feature of the vibration-rotation spectra of
molecules in liquid $^{4}$He nanodroplets is the presence of
rotational fine structure, either quantum state resolved or observed
as rotational contours.  This was first seen in the seminal work of
Hartmann \etal~\cite{Hartmann95} on the spectrum of SF$_6$, but has
now been observed a wide range of other molecular solutes.
Table~\ref{tab:molecules} contains a list of molecules studied to date
with rotational resolution in the IR, at least as known to us.  In almost every
case, the nature of the spectroscopic ro-vibrational structure is
exactly what one expects from the gas phase, i.e., the standard
spectroscopic signatures of a linear, symmetric, or spherical top are
preserved, along with the rotational selection rules based upon the
symmetry of the gas phase transition dipole moment, though often with
dramatically changed spectroscopic constants (see
Section~\ref{sec:rotation} and Figure~\ref{fig:propyne}).  High
temperature spin statistical weights have been observed, indicative of
negligible relaxation between nuclear spin isomers on the time scale
of the experiment \mbox{($\sim 100 \mu$s)}, as first noted by Harms
\etal~\cite{Harms97b} (see also Figure~\ref{fig:propyne}).

Viewed in terms of a classical picture of the instantaneous
configurations of the helium atoms, the above preservation of symmetry
is unexpected in that these instantaneous configurations do not retain
the molecular symmetry.
%The same, however, could also be said of the classical vibrational
%motion of the molecule itself.
In view of the large delocalization of the He atoms, we must rather
consider the symmetry of their ground state many-body wavefunction,
which is the same as the molecular symmetry when the helium is in its
ground state.  Bulk excitations of the helium droplet would lower this
symmetry, but at the temperature of the droplets, liquid-drop-model
calculations indicate that there are no thermally excited
phonons~\cite{Brink90}. Thermal excitations are present in the surface
modes of the droplet, but these appear to couple weakly to the
solvated molecules, though they must ultimately be responsible for the
relaxation of the latter into thermal equilibrium with the droplet.
Thus, we are lead to view the helium not as a classical fluid of
rapidly moving particles, but as a quantum probability density
associated with a bosonic $N$-body wavefunction, which is a smooth and
continuous function of coordinates and reflects the symmetry of the
potential that the molecule creates for the helium.

One potential counterargument of the above model is that linear
molecules will create a cylindrically symmetric ring of helium density
around them, implying that the entire system should act as a symmetric
top, which has an additional rotational degree of freedom.  However,
it is easily shown that the Bose symmetry of the helium eliminates the
low lying $K$ excited states~\cite{Grebenev00a}: states with angular
momentum along the symmetry axis cannot be thermally populated, and
the $K=0$ spectrum of a symmetric top is indistinguishable from that
of a $\Sigma$ state of a linear molecule (see also
Section~\ref{sec:hydrogen}).  This argument alone does not exclude the
possibility of creating $K > 0$ states with modest energy (\mbox{$\sim
  3.2$~K} has been estimated for OCS in He~\cite{Grebenev00a}).
However, in order to create such a ``vortex'' excitation localized in
the first solvation shell, it is necessary to have a nodal surface
separating it from the vortex-free remainder of the helium droplet
\footnote{Nodal surfaces are necessary to separate regions of
  different angular momentum in the helium droplet, in order to
  preserve continuity of the wavefunction.}; such a nodal surface may
come at a steep energy cost, making the creation of high-$K$ modes
even more difficult. It would be interesting to have an estimate for
the excitation energy of such states calculated with the fixed node
diffusion Monte-Carlo (DMC) method.

Roger Miller's group has observed two examples of spectra of highly
polar linear molecules (cyanoacetylene dimer and HCN
trimer~\cite{Nauta99d}) which appear to have Q-branches, and thus
suggest a breakdown of this ``rule'' of preservation of molecular
symmetry.  The origin of the Q-branches in these cases has not yet
been explained conclusively.  It has been proposed that they arise not
from population of higher $K$ states of the helium, but are due to
some mechanism that would create an anisotropic potential for rotation
of the molecules.  One candidate is the anisotropy due to the finite
size of the droplet and the delocalization of the molecule in the
droplet~\cite{Nauta99d} (see Section~\ref{sec:translation}). This
model predicted the wrong droplet size dependence of the Q-branch
intensity unless an \emph{ad hoc} energy term was added to favor
location of the molecule slightly below the helium surface.  Another
proposal~\cite{Miller_Ringberg} is the presence of a vortex in a
minority of droplets, which is expected to strongly align molecules
along it~\cite{Dalfovo99}.  This would also appear to favor Q-branch
intensity in smaller droplets since a larger fraction of pickup
collisions in such droplets would be expected to impart enough angular
momentum to create a vortex (on the order of one $\hbar$ per helium
atom).  We would like to add to the list the anisotropy that arises
from the hydrodynamic coupling of rotation and
translation~\cite{Lehmann99a} (see Section~\ref{sec:translation}),
which is expected to be most important for precisely the extremely
prolate molecules for which this interesting effect has been observed.
Furthermore, this effect should be most important for large droplets,
where the motional averaging of the linear momentum (to which the
angular momentum is coupled) is slowest.  Clearly, these ``odd''
spectra should be subjected to further experimental and theoretical
work since some important new dynamical effect may be revealing itself
by these phenomena.

It is worth noting that none of the ro-vibrational spectra observed in
helium have shown evidence of ``phonon wings''.  Such features, which
appear to the blue of the molecular excitation and involve creation of
one or more helium phonons upon solute excitation, are observed and
often predominate in electronic
spectra~\cite{Stienkemeier95,Hartmann96}.  The strength of the pure
molecular excitation, the zero phonon line (ZPL), is expected to
decrease exponentially with the degree of solvent reorganization upon
molecular excitation, and thus the complete dominance of the ZPL in
the case of vibrational excitation implies that the reorganization of
the solvation density must be negligible relative to the zero point
fluctuations of the helium ground state density.  Perhaps the study of
floppy, large amplitude vibrational modes (see Section~\ref{sec:nh3})
will show evidence of the phonon wings, which could provide valuable
data on solvent reorganization in a quantum liquid.

\subsection{Translational motion of molecules: Where are the molecules?}
\label{sec:translation}

\begin{figure}[bp]
\includegraphics[width=3.375 in]{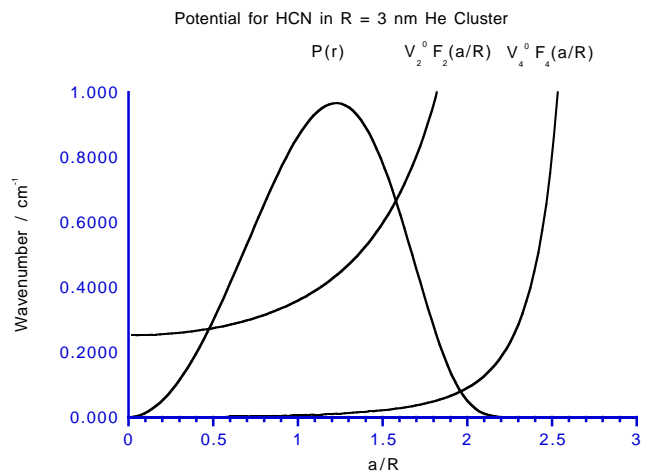}
  \caption{Isotropic (V$_2$) and anisotropic (V$_4$) parts
    of the confining potential for a dopant molecule (HCN) in He
    droplets, arising from the dopant-He long-range interaction.
    $P(r)$ is the radial probability density function for the position
    of the dopant. From Ref.~\onlinecite{Lehmann99a}.}
  \label{fig:fig4}
\end{figure}

The gas-phase-like symmetry of ro-vibrational spectra of molecules
attached to helium droplets strongly suggests that these molecules are
located in a nearly isotropic environment.  This rules out surface
sites since their reduced symmetry would lead to a characteristic
splitting of the $M$ degeneracy, as has been seen for the electronic
spectrum of He$_2^*$ which resides above the surface of
the helium~\cite{Hu00}.  Once a molecule penetrates the helium surface
by more than the first or perhaps second solvation layer, a relatively
flat potential is expected.  Vilesov and Toennies~\cite{Toennies95}
proposed treating the motion of the solvated molecule as that of a
free particle in a spherical box.  However, one can explicitly
calculate the effective potential that pulls the molecule toward the
center of the droplet by summing up the missing long range
interactions outside of the droplet (i.e., taking the molecule in bulk
helium as the reference state).  The resulting potential turns out to
localize molecules well within the droplets (see
Figure~\ref{fig:fig4}), keeping them away from the surface region
where the approximations used to derive the potential are expected to
break down~\cite{Lehmann99a}. The spectrum of thermally populated
energy levels of a particle in this potential resembles more closely
that of a 3-dimensional harmonic oscillator than that of a particle in
a box; the heat capacity arising from this motion is nearly $3
k\si{B}$ and is similar to that of an entire, pure helium nanodroplet
of 3~nm radius.  The effective frequency for vibrational motion in
this potential is \mbox{$\approx 0.5$--1~GHz} in typical cases.  The
potential is proportional to the molecule-helium $C_6$ coefficient
minus the helium-helium $C_6$ coefficient times the net number of
helium atoms displaced by the molecule, i.e., there is a buoyancy
correction due to the change in the net helium-helium interaction as
the molecule is displaced~\cite{Lehmann00}.

For a non-spherically-symmetric molecule, displacement from the
center of the droplet leads to a reduction in site symmetry and a weak
coupling between translational and rotational motion.  The leading
coupling term has a $P_2(\cos \theta)$ dependence on the angle
$\theta$ between the axis of a linear rotor molecule and the
displacement vector from the center of the droplet ($P_n$ are the
Legendre polynomials). If the displacements from the center are
assumed to be static and their distribution given by Boltzmann
weights, this $P_2$ asymmetric field results in the same type of
characteristic splitting of $M$ components of a transition as
discussed above for a surface location, and is easily calculated
analytically.  However, calculations indicate that the expected
splittings are on the order of, or less than, the vibrational
frequency in the 3-D potential, and thus the splittings are predicted
to be at least partially motionally averaged.
Lehmann~\cite{Lehmann99a} has given explicit matrix elements for
calculations in a coupled basis of translational and rotational
angular momenta of the molecule, and presented some sample
calculations.

Another important coupling of translational motion and molecular
rotation is expected to arise from a hydrodynamic effects (see
Section~\ref{sec:movement}).  The translational motion of an ellipsoid in
helium leads to a hydrodynamic contribution to the effective mass,
which depends upon the orientation of the ellipsoid relative to the
linear momentum.  Lehmann~\cite{Lehmann99a} has derived the quantum
Hamiltonian including this coupling term, and the resulting matrix
elements between the translational and rotational angular momenta.
Model calculations on simple systems, such as HCN and OCS, suggest
that the hydrodynamic coupling is more important than the potential
terms mentioned in the previous paragraph.  It is interesting to note
that the hydrodynamic term leads again to the expected $P_2(\cos
\theta)$ induced splittings in the limit of bulk helium, where the
potential asymmetry goes to zero.

The relatively dense manifold of center-of-mass states, coupled to the
rotation of the molecule which changes upon ro-vibrational excitation,
creates a source of inhomogeneous broadening in the spectrum that was
originally not recognized: quantization of the translational motion
should lead to a potentially resolvable fine structure of the spectral
features.  However, it is likely that the broad droplet size
distribution present in all ro-vibrational spectra observed to date
will wash out any details of this fine structure, which is then only
observable as a source of inhomogeneous broadening in the spectrum.

These couplings of the center of mass and rotational motion of the
molecules are believed to make important contributions to the
inhomogeneous lineshapes often observed in rovibrational spectra.
However, calculations done to date have neglected the vibrational
dependence of the interaction of the helium with the molecules.  Any
such model predicts that the lineshape will be independent of
vibrational transition, and that the R($J$) and P($J-1$) transitions
should be mirror images, which is not what is generally observed.
Once vibrationally averaged potentials for different vibrational
states become available, it will be possible to more critically test
the lineshapes predicted by this model.  At present, it must be
admitted we do not have a quantitatively successful theory.

\subsection{Line shapes of rotationally resolved lines}
\label{sec:relax}

\begin{figure}[bp]
\includegraphics{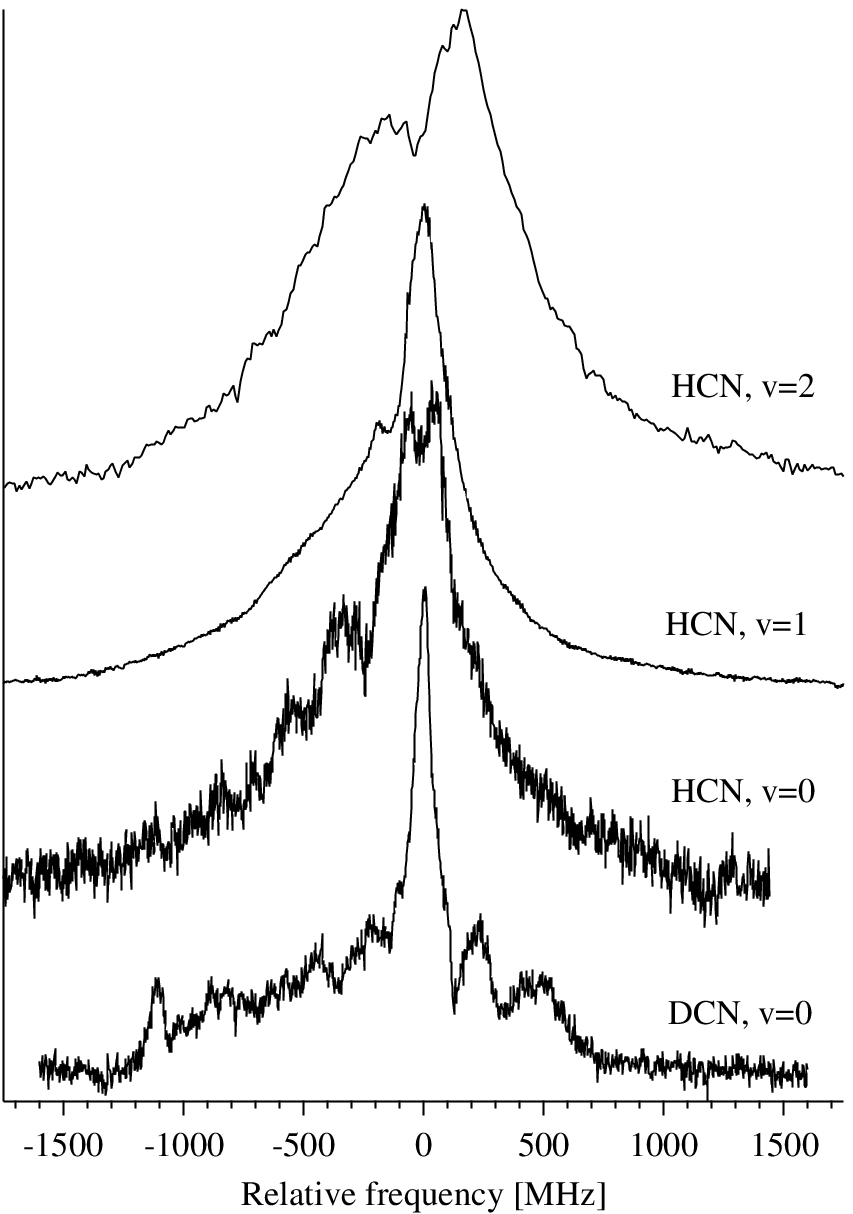}
  \caption{R(0) line for DCN ($\vi{1}=0$) and HCN ($\vi{1}=0,1,2$),
    evidencing the importance of inhomogeneous broadening. These data
    show that vibrational excitation is not very important in
    determining the broadening pattern, whereas the mass of the dopant
    is, suggesting a nontrivial dynamics inside the droplet. The data
    are adapted from Refs.~\onlinecite{Conjusteau00},
    \onlinecite{Conjusteau00}, \onlinecite{Nauta99b},
    \onlinecite{Callegari00c}, respectively.  The zero of the
    frequency was selected to best overlap the transitions.}
  \label{fig:hcn}
\end{figure}

Direct evidence of inhomogeneous broadening is only available for a
few molecules; nevertheless, there are reasons to believe that it is
common to all molecules. One reason is that the ro-vibrational lines
rarely fit to the Lorentzian shapes predicted by simple lifetime
broadening models. This is particularly evident for HCN, for which the
R(0) ($J=0\to 1$) transition has been studied in several vibrational
states, as well as in the deuterated molecule (see
Figure~\ref{fig:hcn}).  For OCS, the lines in the P and R branch have
tails that point towards the band origin and widths that grow with J.
Possible explanations have been discussed by Grebenev
\etal~\cite{Grebenev00a} Available evidence suggests that one can
expect inhomogeneous broadening to contribute a few hundred MHz or
more to the linewidth, equivalent to a lifetime of a few ns.  Often,
rotational and vibrational relaxation is slower than that, yet (with
the notable exception of vibrationally excited HF~\cite{Nauta00c})
still much faster than the 0.1--1~ms time scale of the experiment.

This slow relaxation derives from the large mismatch between the
frequencies of the modes of the molecule and those of the helium bath
(the helium cannot respond on the time scale of the vibrational motion
of the molecule). The maximum energy of an elementary excitation in
bulk $^4$He is about 20~K~\cite{mano86}. As a result, \mbox{V-T} (vibration
to translation) and even most \mbox{V-V} (vibration to vibration) relaxations
require simultaneous creation of many (10--100) phonons/rotons in the
helium. One expects the rate of such processes to be slow, decreasing
exponentially with the number of phonons involved.

\subsubsection{Homogeneously broadened profiles}
\label{sec:relax1}

Exceptions to the above argument are expected in cases of very low
frequency, large amplitude modes, and in cases where there is a
resonance, or near-resonance, of vibrational levels in a molecule and
the optical excitation accesses an upper component of the resonance
polyad. This can be compared to the behavior in the gas phase, where
it is well established that anharmonic resonances allow for rapid
(i.e, approaching the speed expected for rotational relaxation)
\mbox{V-V} collisional relaxation inside the set of coupled levels.

A related phenomenon is the case of molecules which in the gas phase
have a density of states slightly below that required for IVR via
high-order relaxation, such as propyne in the 2\vi{1} mode. In helium,
the effective density of states is raised, as the helium can be
excited with one or more low energy quanta, thus accommodating small
energy mismatches. Molecules with a sufficiently high density of
states to undergo statistical IVR in the gas phase, such as
(CH$_3$)$_3$SiCCH~\cite{Callegari00c}, are at most weakly affected by
the helium.

An interesting case occurs when, in the gas phase, excitation would
cause the molecule to dissociate.  While the helium forms a ``cage''
which may prevents dissociation, it would still accommodate some of the
excitation energy.  As in the case above, one might still be left with
an excited species, but as for determining the lifetime, the initially
excited state would appear to have relaxed. So far this scenario has
only been observed for van der Waals complexes, and the lifetimes are
only moderately decreased relative to their gas phase value. All the
values measured to date are reported in Table~\ref{tab:vibrations}.

\begin{table}[!bp]
% \centering
\caption{Vibrational lifetimes calculated from apparent homogeneous
  linewidths in He nanodroplets (except Ar--HF, measured
  directly). The entry under the header ``Note'' refers to the cause
  of vibrational relaxation; ``tentative'' indicates that vibrational
  relaxation is \emph{assumed} as the dominant cause of broadening.}

  \begin{tabular}{crrcr}\hline\hline
Molecule & Mode & Lifetime/ns & Note & Ref. \\ \hline
HCOOH & \vi{1} (a) & 0.12 & Fermi resonance & \cite{Madeja_preprint} \\
  ''  & \vi{1} (b) & 0.09 &       ''        & \cite{Madeja_preprint} \\
  ''  & \vi{1} (c) & 0.05 &       ''        & \cite{Madeja_preprint} \\
d$_{1}$-HCOOH & \vi{1} & 0.15 & Fermi resonance & \cite{Madeja_preprint} \\
H$^{12}$C$^{12}$CH & \vi{3} & 0.12 & Fermi resonance &
\cite{Nauta_preprint3} \\
H$^{13}$C$^{13}$CH & \vi{3} & $\approx 1$ & Fermi resonance &
\cite{Nauta_preprint3} \\
(HCN)$_{2}$ & \vi{1} & $> 0.64$ & & \cite{Nauta00b} \\
(HCN)$_{2}$ & \vi{2} & 0.15 & dissociation & \cite{Nauta00b} \\
(HF)$_{2}$ & \vi{2} & $>0.53$ & dissociation & \cite{Nauta00b} \\
(HF)$_{2}$ & \vi{2}+\vi{5} & \pt{7}{-3} & tentative & \cite{Nauta00b} \\
(HF)$_{2}$ & \vi{1}+\vi{4} & \pt{3}{-3} & tentative & \cite{Nauta00b} \\
Ar--HF & \vi{1} & \pt{1.5}{6} &  & \cite{Nauta_preprint2} \\
SF$_{6}$ & \vi{} & 0.56 & tentative & \cite{Hartmann99} \\
CH$_{3}$CCH & \vi{1} & 0.16--1.5 & tentative & \cite{Callegari00c} \\
\hline\hline
  \end{tabular}
  \label{tab:vibrations}
\end{table}

All the experimental evidence currently available indicates that the
line profiles are determined by rotational relaxation only when the
energy of one rotational quantum is sufficient to excite bulk modes of
the droplet~\footnote{Particularly compelling is the remarkable
  sharpness of the sub-bands that by selection rules do not
  rotationally relax, observed for HF, (HF)$_{2}$, and HCCH.}. As
the energy must then be near or above the roton minimum of
8.58~K~\cite{ande96}, only small molecules, with large rotational
constants, display this kind of behavior. The observed lines are
Lorentzian, with widths around 1~\wn, reflecting relaxation times in
the 1--10~ps range (see Table~\ref{tab:rotations}).

\begin{table}[!bp]
\centering
\caption{Rotational lifetimes measured in He nanodroplets. The entry
  in the column ``Note'' refers to the droplet modes responsible for rotational
  relaxation.}
  \begin{tabular}{lllll}\hline\hline
Molecule & Levels & Lifetime & Note & Ref. \\
 & or band & (ns) &  &  \\ \hline
HCN/DCN & $J=0$ & $\sim 10$ & surface? & \cite{Conjusteau00} \\
HCCCN & $J=3,4$  & 2--20 & '' & \cite{Reinhard99} \\
  ''  & $J=2,4$ & 10--100 & '' & \cite{Callegari00b} \\
HF & \vi{1} & \pt{12}{-3} & bulk & \cite{Nauta00c} \\
(HF)$_{2}$ & \vi{1} & \pt{2.7}{-3} & '' & \cite{Nauta00b} \\
\hline\hline
  \end{tabular}
  \label{tab:rotations}
\end{table}

\subsubsection{Indirect measurements: Saturation, double resonance}
\label{sec:relax2}

For large molecules, the dissipation channel into bulk helium modes is
closed for thermally populated rotational states, and relaxation must
occur either by \mbox{R-T} (rotation-translation) transfer (for which
Franck-Condon factors are unfavorable~\cite{Lehmann98}) or by coupling
to the softer surface modes of the droplet. Because long-range forces
confine the molecule to spend most of its time away from the surface,
the latter coupling is weak, and relaxation times are correspondingly
extended into the nanosecond range (see Table~\ref{tab:rotations}). In
fact, the coupling to the surface modes has been estimated to be so
weak~\cite{Lehmann99a} that it is an outstanding question how the
molecules can come to equilibrium with such a grainy heat bath. Lines
are then inhomogeneously broadened, but homogeneous widths can still
be estimated from saturation and double resonance measurements.

The first such measurement, the MW-MW experiment on
HCCCN~\cite{Reinhard99}, faced a set of apparently contradictory findings: the
single resonance saturation data demonstrated that the strength of
rotational lines was proportional to the square root of the applied
power over a wide range, the classic signature of saturation of a line
dominated by inhomogeneous broadening~\cite{Demtroeder96}. As the only
previously considered source of inhomogeneous broadening was the
spread of droplet sizes, a standard analysis based on the assumption
of ``static'' broadening predicted that double resonance measurements
would directly reveal the homogeneous component of the line via
hole-burning with the the pump field, i.e., a Bennet hole and hill in
the absorption of the probe~\cite{Demtroeder96}, with a width of no
more than 150~MHz. It was instead found that the linewidths of the
transitions affected by the pump beam were comparable to those
measured in single resonance (1~GHz). The contradiction was solved by
ascribing inhomogeneous broadening to the spectral diffusion of the
dopant over a set of quantum states (most likely different
translational states within the droplet; see
Section~\ref{sec:translation}). Initially it was assumed that such
diffusion occurred on a time scale faster than rotational population
relaxation. Upon further analysis, it was realized that this would
have corresponded to a much higher saturation power than was actually
observed, and it was proposed that the two phenomena are occurring on
the same time scale (tens of ns)~\cite{Callegari00b}.  Despite the use
of a simplified model, and the tentative value of the parameters
obtained from it, the conclusion that spectral diffusion and
rotational relaxation both occur because of the motion of the dopant
inside the droplet appears to be reasonable; indeed the time scale for
these is in the ns range.

\begin{figure}[bp]
\includegraphics[width=3.375 in]{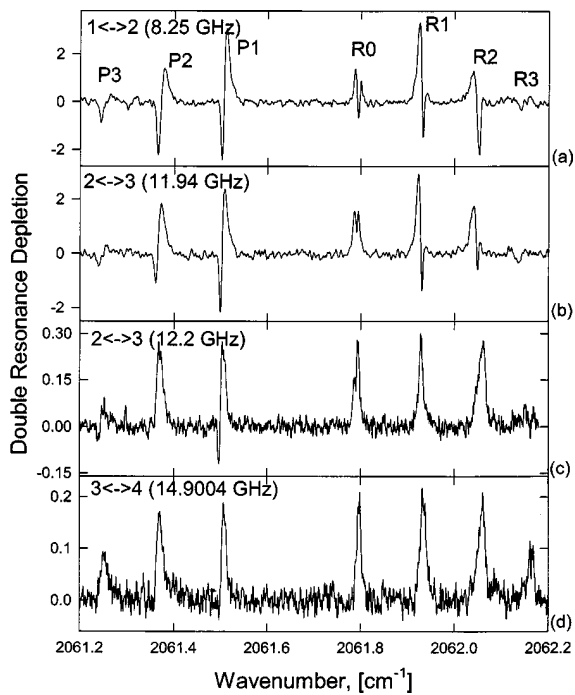}
  \caption{Double resonance spectra of OCS in He droplets. The levels
    coupled by the MW ``pump'' and the frequency of the transition are
    indicated in the upper left corner of each panel. In contrast to
    what is commonly observed in gas phase experiments, note how
    \textit{all} the levels are affected by the MW ``pump'', and how
    the total signal does not sum to zero. From
    Ref.~\onlinecite{Grebenev00b}.}
  \label{fig:ocs}
\end{figure}

MW-IR experiments performed on HCCCN~\cite{Callegari00b} and
OCS~\cite{Grebenev00b} confirmed the above picture: double resonance
lines span the whole inhomogeneous width. A reproducible substructure
of 5 peaks, about 50~MHz wide is observed for the $J = 2 \rightarrow 3$
rotational transition of OCS,
suggestive of a $2J+1$ splitting, but no pattern is resolved for other
lines, thus making a definite assignment not possible.  The main
finding, common to both experiments, and therefore probably of general
nature, is that the populations of \textit{all} the rotational levels
are altered by the MW ``pump'' (Figure~\ref{fig:ocs}).

In gas phase DR experiments, such as the classic IR-MW studies by
Oka~\cite{Oka73}, one can often observe collisionally-induced four
level DR signals, but they are typically considerably weaker than the
collisionless features.  In helium, the relaxation induced signals
have been found to be even stronger than the ``direct'' signals in
some cases.  There also was a strong dependence of the DR signals on
the source conditions, which suggests that the relaxation dynamics are
strongly dependent upon droplet size, which affects the density of
states of the thermal bath that the molecule is coupled to.  For
HCCCN, the DR spectra dramatically simplify for the largest droplets,
which has suggested rotational relaxation is in the ``strong
collision'' limit for these droplets, i.e., that the relative
populations produced by relaxation are independent of the initially
populated state.
% For OCS, there appeared to be a modest systematic sharpening of the IR
% transitions when irradiated by the MW radiation, but not consistent
% with the expectation of a hole burning model.  Further, the R(0)
% transition appears to develop a substructure hidden in the pure IR
% spectrum.

A curious feature observed for both  HCCCN and OCS is that there
appears to be a lack of conservation of population in the DR
experiment.  For DR signals that arise from population transfer, the
integral of the DR signal over the rotation structure of the IR band
should be zero since the integral cross section for IR absorption is
independent of the lower state.  This apparent lack of population
conservation must reflect some ``artifact'' in the experiments, but
the ones considered do not appear consistent with the observations; in
particular, the sign of the total signal is opposite for the two
molecules.  This effect is currently unexplained and may ultimately
reveal some interesting novel dynamics for these systems.

\subsection{Molecular rotations}
\label{sec:rotation}

\subsubsection{Changes in rotational constants: How is the helium moving?}
\label{sec:movement}

An important aspect of the rotational structure of solvated molecules
is that in most cases the effective rotational constants are
substantially reduced from their gas phase values.
%Figure X contains a plot of the
%ratio of the cluster to gas phase $B$ value, as a function of the gas phase
%$B$ value.  While there is considerable scatter in the data, it is
%evident that
For molecules with rotational constants much larger than 1~\wn, the
change on going into helium is small, while for heavier molecules,
the rotational constant is typically reduced by a factor of 3--4 upon
helium solvation.  Clearly, when the molecules rotate in helium, at
least some of the helium must also move, creating kinetic energy and
contributing to the effective moment of inertia of the molecular
rotation.  Several models have been put forward to describe the nature
of the helium motion.

The first model that was proposed~\cite{Hartmann95,Grebenev00a}
assumed that a specific number of helium atoms were rotating rigidly
with the molecule, directly contributing their moment of inertia to
that of the molecule.  This model appeared to work for both SF$_6$ and
OCS, in that chemically reasonable positions and numbers of rigidly
attached helium atoms were consistent with the experimentally observed
changes in moment of inertia ($\Delta I$).  For OCS, the authors
called this the ``donut model'' since it assumes that a ring of 6~He
atoms is localized in the potential minimum locus around the OCS
molecule.  This very simple model has two problems.  First, when one
tries to extend it to lighter molecules such as NH$_3$ or HCN, even a single
rigidly attached He atom would increase the moment of inertia by
nearly an order of magnitude more than what is observed
experimentally.  Second, more recent results for the \mbox{OCS--H$_2$}
complex~\cite{Grebenev01} in helium are clearly inconsistent with the
``donut model'' for OCS\@.  In this work, it was found by isotopic
substitution that a single H$_2$, HD, or D$_2$ molecule will replace
one of the helium atoms in the ``donut''.  Yet, extrapolation to zero
mass of the isotopomer gives a moment of inertia of the
``donut with a bit taken out'' considerably larger than for the OCS in
a pure Helium droplet, i.e. with a complete ``donut''.

A more sophisticated ``two fluid'' model for the motion of the helium
was introduced by Grebenev \etal~\cite{Grebenev98}.  They proposed
that the density of helium around the molecule be partitioned into
spatially dependent normal and superfluid fractions, with the normal
fraction having large values only in the first solvation layer.  In
analogy with the classical Andronikashvili experiment (where the
normal fluid fraction of bulk helium follows the rotation of a stack
of plates, contributing to their effective moment of
inertia~\cite{Andronikashvili46}), these authors propose that the
normal fluid fraction rotates rigidly with the molecule, contributing
the classical moment of inertia of the normal fluid mass density.
This model was incomplete, however, in that the authors provided no
definition of the spatially dependent normal fluid density that
appears in the expression for the increase in moment of inertia.

Kwon and Whaley proposed an \emph{ansatz} for a local superfluid
density~\cite{Kwon99b}, based upon an extension of Ceperley's global
estimator for the bulk superfluid density from path integral Monte
Carlo (PIMC) calculations~\cite{Ceperley95}.  They took the difference
in total density and superfluid density to be a nonsuperfluid density
that rotates rigidly with the molecule, assuming the molecule-He
interaction is sufficiently anisotropic.  The term ``nonsuperfluid
density'' is used instead of ``normal fluid density'' because this
density is not related to the normal fluid density produced by a
``gas'' of elementary excitations, as in bulk helium experiments such
as the Andronikashvili experiment.
The inertial response of the helium to rotation is a second rank tensor,
involving both the axis of rotation and the direction of the response,
which is measured in PIMC by a projected area of closed Feynman
paths~\cite{Ceperley95}. The Kwon-Whaley estimator produces a scalar
superfluid density, and thus has the wrong symmetry to describe the
helium inertial response~\cite{Ceperley_Ringberg,Draeger_thesis}.  This two
fluid
model has, however, made predictions in excellent agreement with experiment for
the cases of SF$_6$ and OCS in helium.
Draeger and Ceperley have proposed a new local superfluid density
estimator that has the correct tensor properties and gives, when
spatially averaged, the correct value for the global superfluid
fraction~\cite{Draeger_thesis}.

An alternative, quantum hydrodynamic model has also been developed for
the helium contribution to the moment of inertia of molecules in
helium.  It is well known that linear motion of a rigid body through
an ideal (aviscous and irrotational flow) fluid generates motion in
the fluid that contributes to the effective mass for translation.  For
example, for a sphere, the effective mass increase is exactly one half
of the mass of the displaced liquid~\cite{Lamb1895}.  Similarly,
rotation of an ellipsoid generates helium kinetic energy proportional
to the square of the angular velocity, thus contributing to the
effective moment of inertia.  The possible inclusion of such a
``superfluid hydrodynamic'' term was already recognized by Grebenev
\etal~\cite{Grebenev98}.  However, attempts to estimate this term,
using classical expressions for the moment of inertia of an ellipsoid
in a uniform fluid of the density of bulk helium, gave results which
were only a small fraction of the experimental
values~\cite{Grebenev98,Lehmann99a}.  On the other hand, as the
density around a molecule is highly nonuniform, Callegari
\etal~\cite{Callegari99} have developed a quantum hydrodynamic model
that properly takes into account the anisotropy of the density, and
have obtained moments of inertia in good agreement with experiments
for a number of molecules.

\begin{figure}[bp]
\includegraphics{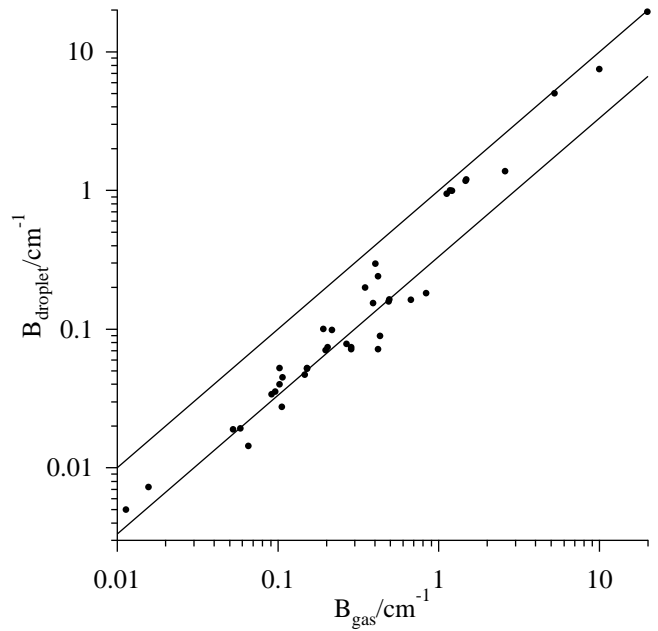}
  \caption{Plot of the rotational constants (gas phase versus helium
    droplet) reported in Table~\ref{tab:molecules}. The two lines are
    the functions $y=x$ and $y=x/3$ respectively. Note the different
    trend above and below $B_{\mathrm{gas}}=1$~\wn, indicative of a
    threshold for adiabatic following.}
  \label{fig:bplot}
\end{figure}

If one adiabatically separates the helium motion from the rotation of
the molecule, then one expects the helium density in the rotating
frame of the molecule to be constant and equal to that around a static
molecule~\cite{Lee00,Callegari99}.  This is called the adiabatic
following approximation and, on physical grounds, should be accurate
when the minimum rotational spacing, $2B$, is well below the energy of
the roton, which characterizes the helium density excitation energy on
the atomic level (in the two fluid model, Kwon \etal\ have introduced
a distinct definition of adiabatic following~\cite{Kwon99b,Kwon00}).
If one further assumes that the fluid velocity field generated by this
time dependent helium density in the laboratory frame is irrotational
(which Kelvin proved to be the lowest kinetic energy
solution~\cite{Milne96}), then one obtains from the equation of
continuity a second order differential equation for the velocity
potential (the instantaneous laboratory frame velocity is the negative
gradient of this velocity potential).  The helium contribution to the
moment of inertia is then calculated directly as an integral over the
velocity potential.  In applying this method to molecules of
cylindrical symmetry, the helium density was calculated using a
density functional (DF) approach~\cite{OrsayParis,OrsayTrento} which,
instead of searching for the ground state of the helium many-body
problem, finds the solution of a one-particle problem with a nonlocal
self interaction, whose density distribution, if the correct
functional was used, would exactly match the density of the many-body
wavefunction.  It takes orders of magnitude less computational effort
to solve for the helium density by DF methods than by Monte Carlo
methods, and the results have been found to be in good agreement for
cases where the molecule-helium interaction potentials are not too
deep~\cite{dalf94}.

An alternative derivation~\cite{Lehmann00_unpublished} of the quantum
hydrodynamic approach is to write the many-body helium wavefunction in
a rotating ``external'' molecular potential as that of the
instantaneous static ground state wavefunction times a single particle
phase function.  Variational optimization of the energy with respect
to this phase function gives the same hydrodynamic equation as for an
ideal classical fluid with the adiabatic following approximation.
Within a Hartree approximation for the helium wavefunction (which DF
theory implicitly invokes), the hydrodynamic treatment gives the exact
wavefunction correction to first order in angular velocity and thus
the energy to second order.  Time
reversal symmetry implies that changes in the helium density can only depend on
even powers of the angular velocity, and will therefore contribute only to
the centrifugal distortion constant, not to the rotational constant.
It has been known since the dawn of quantum
mechanics~\cite{Madelung26,deBroglie26} that quantum mechanics can be
formulated \emph{exactly} as a hydrodynamic problem in configuration
space.  The one body approximate hydrodynamic approach used for these
calculations is equivalent to many calculations that have been done in
bulk helium, including the structure of the core of a
vortex~\cite{Ortiz95}, which has a helium density structure on a
length scale smaller than that induced by the solvation of most
molecules.

The quantum hydrodynamic model has been applied to a number molecules,
including HCN, HCCH, HCCCH$_3$, OCS, HCCCN, and HCN
dimer~\cite{Callegari99}.  For the last four, for which the adiabatic
following approximation is expected to hold, the predicted increase in
moment of inertia was between 103-130 \% of the observed values, as
good as could be expected given the uncertainties in the helium
solvation densities.  For HCN and HCCH, which have small moments
inertia, the hydrodynamic calculations significantly overestimate the
observed increase which is taken as evidence of the breakdown of the
adiabatic following (see below).  Kwon \etal~\cite{Kwon00} have
reported that the hydrodynamic model predicts for SF$_6$ an increase
in the moment of inertia of only 6(9)\% of the experimentally observed
value when their calculated superfluid (total) helium solvation
density was used.  Similar hydrodynamic calculations have been done by
Lehmann and Callegari~\cite{Lehmann02a} using the lowest four tensor
components in the expansion of the helium density in octahedral
harmonics.  These authors find that when the total helium density is
used, the hydrodynamic contribution to the moment of inertia was 55\%
of the observed value, and this value is likely to increase if the
full anisotropy of the helium density is included.  The reason for
the discrepancy in the two reported values~\cite{Kwon00,Lehmann02a}
for the hydrodynamic prediction for SF$_6$ has not yet been resolved.

Despite their apparent differences, the ``two fluid'' and the hydrodynamic
approaches to the superfluid helium rotation problem are related.
They are both based on a linear response of the helium, and thus not
applicable when the adiabatic separation of molecular rotation and
helium motion breaks down.  Furthermore, the ``nonsuperfluid'' helium
accounts for much of the angular anisotropy of the helium density,
which is the source term in the quantum hydrodynamic treatment.

Clearly, it would be desirable to develop a theoretical method to
directly calculate the rotational constant of molecules in superfluid
helium without making the dynamical approximations of either the ``two
fluid'' or quantum hydrodynamic approaches.  It would also be
desirable to directly calculate relaxation rates for excited
rotational states as well as their energies.  Unfortunately, this is
still largely an unsolved problem.  Blume \etal~\cite{Blume99} have
used an excitation energy estimator based upon the rate of decay of an
approximate excited state wavefunction under imaginary time
propagation (POITSE method).  Unfortunately, only studies of very
small clusters have been published to date with this method, and thus
its relevance to rotational dynamics in helium nanodroplets is not
established.  Further, the initial state had the same rotor based
nodal surfaces and thus the accuracy of this method may also depend
upon an approximate adiabatic separation of the rotational motion from
that of the helium.  If this approximation breaks down, the initial
state will contain sizable contributions from several excited states
and the imaginary time propagation will contain sizable contributions
from many exponential decays.  Lee \etal~\cite{Lee00} have calculated
the rotational constant of \mbox{SF$_6$--He$_N$} ($N = 1$--20) by a
fixed-node diffusion Monte Carlo method.  In these calculations, the
helium and the molecule moved in the same coordinate system, a
condition that is particularly easy for a spherical top molecule, but
harder in general.  Calculations for the rotational spacings of SF$_6$
were in quantitative agreement with experiment for $N \ge 8$, as in
the simple model proposed earlier by Hartmann \etal~\cite{Hartmann95}
The fixed node approximation used, based upon the nodes of the
rotational wavefunction of the isolated molecule, was justified by
precisely the same type of adiabatic separation of time scales that is
used for adiabatic following in the quantum hydrodynamic
approach~\cite{Lee00}.  Recent work on a model system~\cite{Lehmann01}
has found that the hydrodynamic and this molecular rotor based fixed
node approximation are closely related, and break down at about the
same point when the rotational constant of the molecule is increased.
To date, the diffusion Monte Carlo calculations have been published
only for SF$_6$, so the accuracy of the method in general is hard to
judge, though it is likely the most accurate of the presently
available methods.

\subsubsection{Breakdown of helium adiabatic following}

As it was mentioned in the beginning of Section~\ref{sec:movement},
``light'' molecules with large gas phase $B$ values appear to have
little helium contributions to their moments of inertia.  It is
certainly the case that these molecules, by their nature, have
relatively weakly anisotropic interaction potentials with the helium,
and thus a small increased moment of inertia is expected.  However,
because they also have very small molecular moments of inertia, even a
small helium contribution would be expected to lead to a substantial
fractional change in the value of $B$\@.  This suggests that something
else is going on.  The models discussed above invoke ``adiabatic
following'' of the helium with the molecular rotation, though the
precise meaning of that phrase is different for the quantum
hydrodynamic and the ``two fluid'' models.  It is expected that this
adiabatic separation of the time scales of molecular rotation and
helium motion will break down when the rotational spacing becomes
comparable to the energy required to create a helium excitation with a
wavelength on the order of the solvation structure around the
molecule, and this will be similar to the energy of the helium roton,
8.58~K \cite{ande96}.  This suggests a rotational constant of
\mbox{$\approx 4.3$~K} as the point where substantial breakdown of the
adiabatic following would be expected (see Figure~\ref{fig:bplot}).
Lee \etal~\cite{Lee00} studied adiabatic following breakdown
theoretically by changing the rotational constant of SF$_6$ in their
fixed node DMC calculations of small SF$_6$-He$_{\rm N}$ clusters.
They found that the angular anisotropy of the helium in the molecular
frame and the helium contribution to the effective moment of inertia
both decreased substantially when the SF$_6$ rotational constant used
in the calculation was increased by a factor of ten, from $0.1
\rightarrow 1 {\rm cm}^{-1}$.  HCN ($B=2.13$~K \cite{maki74}) is one
of the few molecules that have been studied which is near this
threshold.  The breakdown of adiabatic following for HCN has been
established experimentally, by observation of the change in the helium
contribution to the effective moment of inertia by isotopic
substitution of HCN to DCN\@.  Compared to DCN, it was found that the
somewhat lighter and thus 23\% faster HCN produced a 9\% decrease in
the helium contribution to the moment of inertia \cite{Conjusteau00}.
It would be highly desirable to study other molecular systems in this
important intermediate region.

Density-functional computations can reproduce the breakdown of
adiabatic following to some degree by explicitly moving the density
computation from the inertial lab frame to the frame rotating with the
dopant molecule. In terms of the helium and total angular momentum
operators $\vec{L}$ and $\vec{J}$, respectively, the rotating-frame
rotational energy of the dopant molecule is $B(\vec{J}-\vec{L})^2=B
J^2+B L^2-2 B \vec{J} \cdot \vec{L}$ \cite{huts91}. In the rotational
ground state, $\langle J^2 \rangle=0$, and $\langle \vec{L} \rangle=0$
because of time reversal symmetry.  Therefore, the only change in
going to the rotating frame is the appearance of a term $B L^2$ in the
helium density functional. For large values of $B$, this term
increases the angular smoothness of the ground state helium density,
which decreases the helium moment of inertia in the hydrodynamic
theory; this provides a scale to gauge the breakdown of adiabatic
following.  Work is in progress in our group to quantitatively assess
this approach.
%Calculations including this new term, combined with the hydrodynamic
%model, have given results in excellent agreement with experiment for
%HCN and DCN: $\Delta I\si{HCN}\se{DFT} = 2.8$~u\AA$^2$ and $\Delta
%I\si{DCN}\se{DFT} = 3.05$~u\AA$^2$~\cite{Schmied00}, to be compared to
%the experimental values of $\Delta I\si{HCN}\se{exp} = 2.6$~u\AA$^2$
%and $\Delta I\si{DCN}\se{exp} = 2.9$~u\AA$^2$~\cite{Conjusteau00}.

\subsubsection{Giant centrifugal distortion constants}

Another interesting generic feature that has been observed in
rotational spectra in helium droplets is a dramatic increase in the
centrifugal distortion constants required to fit the spectra.  Values
of $D/B \sim 10^{-4}$ have been observed for heavy molecules, three to
four orders of magnitude larger than for the same molecules in the gas
phase.  It has been suggested that these large distortion constants
reflect weakly bound helium atoms that are easily displaced by the
centrifugal potential, but a careful attempt to estimate this effect,
for OCS, gave a predicted $D$ value 50 times smaller than is observed
experimentally~\cite{Grebenev00a}.  If one uses the standard scaling
$D \sim 4 B^3/\omega^2$, one must invoke an effective vibrational
wavenumber on the order of $1$~\wn\ in order to reproduce the observed
$D$ values, much smaller than what is physically reasonable
considering the force constants of the helium-molecule potentials.
Given the above discussion of the breakdown of adiabatic following
with increasing rotation rate, one can argue that the effective $D$
value should be negative, the opposite sign from what has been
observed!  This is because we would expect the effective $B$ value to
increase, going towards the gas phase value, as the rotational
velocity increases, making it harder for the helium to follow.

At present, there is only one calculation that has reproduced the
correct size of the distortion constant: a toy model~\cite{Lehmann01}
of a rigid, planar ring of $N$ He atoms interacting with a rotating
molecule.  What this calculation clearly demonstrated is that opposite
behaviors are found depending on how the rotational velocity is
increased: by decreasing the moment of inertia of the molecular rotor
or by spinning up a fixed rotor to higher $J$ values.  In the former
case, the helium follows more poorly, consistent with the above
discussion. In the latter case, the He anisotropy and following
increase until a resonance between the molecular and helium ring
rotations occurs.  These results are consistent with general
conclusions previously drawn by Leggett in his analysis of the
``rotating bucket'' experiment for bulk liquid
helium~\cite{Leggett73}.

\subsubsection{Vibrational dependence of rotational constants}

As in the gas phase, the effective rotational constants change upon
vibrational excitation.  The size of the observed vibration-rotation
coupling is much larger in helium than that expected based upon the
gas phase change in $B$\@. For example, for HCCCN the total increase in
moment of inertia upon vibrational excitation changes from
0.35~u\AA$^2$ in the gas phase to 9.2~u\AA$^2$ in helium, and for
CH$_3$CCH it changes from 0.32~u\AA$^2$ to
8.5~u\AA$^2$~\cite{Callegari00c}.  This suggests that most of the
effect is due to a change in the helium contribution to the effective
moment of inertia.  As expected, the excited vibrational states have
increased helium contributions, which suggests increased density
and/or anisotropy around the vibrationally excited state.  As pointed
out above, the absence of phonon wings to the spectra indicate that
these changes must be small.  There has yet to be published a quantitative
attempt to calculate this change in solvation structure, which
requires knowledge of the vibrational dependence of the molecule-He
interaction potential, which are not yet generally available.
Recently, there has been important progress made in the calculation of
the intermolecular-mode dependence of the interaction
potentials~\cite{Szalewicz01}, which would allow such an estimate to
be made.

\subsection{Shift in band origins}
\label{sec:shifts}

An extremely attractive feature of ro-vibrational spectroscopy in
helium is that solvent shifts of the vibrational origins are generally
quite small, typically less than $0.1\%$.  This can be contrasted with
spectroscopy in the traditional matrix environments of Ne or Ar, for
which the shifts are usually much larger. As an example, for HF the
matrix shifts are 0.067\%, 0.24\% and 1.1\%,
respectively~\cite{Huisken97}.  The small shifts in helium are due in
large part to its low density and polarizability (even when compared
to other rare gases).  However, existing evidence suggests that there
is often a fortuitous cancellation of effects leading to smaller
shifts than one could otherwise expect.

In considering the shift in band origin, one must distinguish between
the \emph{vertical} and \emph{adiabatic} excitation energies.  The
former refers to a transition within a frozen helium solvation
density, while the latter considers the optimized helium solvation
density in each vibrational state.  As discussed above, the observed
vibrational dependence of the effective rotational constants is
largely a result of such a change in helium solvation.  The peak of
the zero phonon line should provide an experimental estimate for the
adiabatic excitation energy while the vertical excitation will be at
the `center of gravity' of the entire transition, including the phonon
wing.  The difference between vertical and adiabatic excitation
energies reflects the helium reorganizational energy during
vibrational excitation.  Given the lack of observed ``phonon side
bands'' in vibrational spectra, one can anticipate that this
reorganizational energy is small, but not necessarily small compared
to the typically observed helium induced vibrational frequency shift.
If a $\approx 10$\,cm$^{-1}$ wide phonon wing contains 50\% of the
integrated intensity of a transition, it would cause a $\approx
5$\,cm$^{-1}$ shift between the adiabatic and vertical excitation
wavenumber, but this phonon wing would have a peak absorption strength
less than 1\% of that of the zero phonon line (assuming a linewidth of
1-3 GHz), and thus could be easily lost in any low frequency noise in
the spectrum.  The vertical excitation energy, in contrast, is much
easier to estimate theoretically, since it does not require
computation of the change in helium-helium interactions.  As an
example, Blume \etal~\cite{Blume96} have calculated the vertical
excitation wavenumber of HF in helium by averaging the potential
difference for He interacting with v=0 and 1 states of HF over the
ground state wavefunction of the system.

The change in the helium-molecule interaction potential upon
vibrational excitation reflects several different physical components.
Based upon the natural separation in time scales between
intramolecular and helium solvation motions, an adiabatic separation
is appropriate and one can consider a vibrationally averaged
molecule-helium interaction potential for each vibrational
state~\cite{Szalewicz01}.  If we consider excitation of a normal mode
$k$ in a molecule with dimensionless normal coordinates $q_i$, then in
the harmonic approximation the He-molecule potential in the excited
state of the normal vibration $\nu_k$ can be written as
\begin{equation*}
V_{\rm eff} = V_{\rm g} + \sum_s \left( \frac{\partial V}{\partial q_s} \right)
\langle q_s \rangle + \frac{1}{2}\left( \frac{\partial^2 V}{\partial
q_k^2} \right) \langle q_k^2 \rangle,
\end{equation*}
where $V$ is the full interaction potential, including intramolecular
coordinates, and the derivatives are evaluated at the equilibrium
structure of the isolated molecule.  The sum is restricted to totally
symmetric vibrational modes, and $\langle q_s \rangle$ is
proportional to the cubic anharmonic coupling $\phi_{skk}$ between the
excited mode $k$ and the symmetric mode $s$.  Given the small changes
in vibrational wavenumber upon solvation by helium, the harmonic
approximation is expected to be adequate for rigid molecules.

The long range interaction between He and the molecular dopant is
dominated by the isotropic and $P_2(\cos \theta)\,R^{-6}$ terms.  One
term contributing to the vibrational dependence of this term is the
polarization stabilization of the dipole induced by vibrational
motion, even in nonpolar molecules such as SF$_6$~\cite{Eichenauer88}.
This term leads to a shift proportional to the IR strength of the
vibrational mode.  Integrating this term over an estimate for the
helium solvation density gave a shift for SF$_6$~\cite{Kwon96} of
about one half the observed value of
$-1.6$~\wn~\cite{Goyal92b,Hartmann95} (such an estimate is for the
vertical, not adiabatic shift).  A sizable fraction of the shift comes
from the first solvation shell, for which the use of the long range
$R^{-6}$ terms is not expected to be complete, though there may be a
significant cancellation of errors between the neglect of higher order
terms and damping of the dispersion series~\cite{Douketis82}.  For
molecules with a permanent dipole moment, one also needs to consider
the changes in the vibrationally averaged dipole moment.  While these
changes are usually smaller than the transition moments, at least for
reasonably strong IR transitions, one gets an ``interference'' with
the permanent dipole moment (since the induction energy depends on
$\langle \mu^2 \rangle$), which can magnify the effects of a small
absolute change in the dipole moment upon vibrational excitation.  For
excitation of the overtone of the \mbox{C-H} stretching mode in
HCN~\cite{Callegari00c}, the transition dipole and change in
vibrationally averaged dipole moment lead to estimated shifts of 0.14
and 1.67~\wn, respectively.  The long range interaction between He and
molecular dopants is typically dominated by dispersion interactions,
not the induction terms.  One can estimate the change in the induction
terms from changes in the polarizability of the molecule upon
vibrational excitation.  For HCN, this gives an estimate of 4.98~\wn\
for the shift, substantially larger than the observed value.

All of the terms discussed above are long range and lead to red
shifts.  They certainly are expected to dominate the droplet size
dependence of the shift.  Putting the solute in the center of the
droplet and integrating the missing (compared to the bulk)
interaction, one finds that the shift from bulk or infinite droplet
size limit will be dominated by the $R^{-6}$ terms and will scale as
$N^{-1}$, where $N$ is the number of helium atoms in the
droplet~\cite{Jortner92}.
Thus, a red shift towards the limiting bulk value is expected for
large droplets, independent of whether the overall shift is red or
blue.  If one assumes that the droplet size distribution follows the
log-normal distribution, then the inhomogeneous distribution of
droplet sizes will also contribute a log-normal distribution of shifts
away from the bulk value.  This allows the $N$ dependence of the width
and shift for smaller droplets (where the inhomogeneous cluster size
distribution contributes most to the spectral shape) to be used to
estimate the inhomogeneous size dispersion contribution to the lineshape of
larger
droplets.

Counteracting the above discussed increases in long range He
attraction upon vibrational excitation (and higher order multipole
contributions that have not yet been estimated) is an expected blue
shift from the ``crowding'' interaction.  Such an effect is due to the
fact that the potential for a molecular oscillator will be in part
stiffened due to the interaction with the first solvent shell, where
repulsive forces are important.  In a highly simplified harmonic
picture of a linear \mbox{X-H--He} unit, the \mbox{X-H} stretch normal
mode force constant will be increased by the \mbox{H-He} force
constant.  This leads to a blue shift of the vibrational frequencies
that should be most important for vibrational motion dominated by
exposed atoms and/or involving low frequency, large amplitude motions.
Examples of the later include the torsional mode of
glyoxal~\cite{Hartmann96a} and the butterfly mode of
pentacene~\cite{Hartmann01}, both in the first excited electronic
states. At fixed helium solvation structure, one would expect that the
blue shift of a chemically unique vibrational motion, such as the
\mbox{C-H} stretching mode of terminal acetylene compounds, would be
approximately conserved. However, helium density functional theory
calculations made by our group have shown that helium solvation
density around this terminal \mbox{C-H} bond is a strong function of
the rest of the molecule, which thus far has prevented the development
of a quantitatively predictive theory of the expected blue shift.

The spectrum of HCN dimer provides some evidence for the importance of
the blue shift induced by the local structure.  For the hydrogen bound
\mbox{C-H} stretch, Nauta and Miller~\cite{Nauta99c} have observed a
larger gas-to-helium-droplet redshift compared to the same quantity
for the HCN monomer.  They interpreted this as arising from an
increasingly linear and thus strengthened hydrogen bond when HCN dimer
is solvated in helium.  However, this effect can be also interpreted
as arising from a substantial reduction of the helium induced blue
shift upon intermolecular complex formation since the hydrogen bonded
H atom no longer pushes directly on the helium solvent.

At present, we lack an effective theory to use the observed solvent
induced shifts to learn about the solvation structure around the
chromophore molecule or to reliably predict the observed shifts, even
their sign, relative to the gas phase.  However, since the observed
shifts appear to be small, we can safely neglect this complication in
most cases, in particular when using the predictions of quantum
chemistry to help assign novel spectra observed in liquid helium.

\section{Special topics}
\label{sec:special}

\subsection{Synthesis of nonequilibrium structures}
\label{sec:noneq}

A very promising application of HENDI spectroscopy which has been
recently and dramatically demonstrated by Nauta and
Miller~\cite{Nauta99a,Nauta99d,Nauta00a} relates to the possibility of
using the very cold droplet environment to synthesize highly
non-equilibrium species that could not be prepared in any other known
way. In analogy to the preparation of van der Waals clusters of spin
polarized alkali atoms demonstrated in Princeton in 1996 (see the
following review by Stienkemeier and
Vilesov~\cite{Stienkemeier01_here} and
Refs.~\onlinecite{Toennies01,Higgins96,Higgins00}), the essential
ingredient in these syntheses is, in addition to the low temperature
of the droplets, the rapid cooling action that the superfluid helium
exerts on the molecular partners at the time of cluster formation.
Consider the example of the preparation of linear chains of
HCN~\cite{Nauta99a}: such chains with more than three monomer units
have not been observed even in the cold environment of a supersonic
jet expansion, where cyclic or other nonpolar structures are instead
prepared.  When an HCN molecule collides with a cluster that already
hosts another molecule in its interior, the long range dipole-dipole
forces align the two dipoles at distances where no other force is
acting between the two molecules. The rapid rotational relaxation
necessary for such an alignment is probably provided during multiple
collisions with the droplet surface, where coupling to ripplons is
enhanced.  After the two dipoles have been aligned, the molecules are
drawn together by their mutual attraction; in contrast to a gas phase
coagulation, however, their potential energy is not completely
transformed into kinetic energy because of the steady cooling action
of the ``solvent''. Consider that a potential energy of 200~K (i.e.,
\pt{2.8}{-21}~J) would accelerate a molecule of mass 27~u to a
velocity of 350~m/s, which is well above the critical velocity of
60~m/s~\cite{Wilks87} above which a foreign body will rapidly
dissipate energy in superfluid helium. This dissipative cooling during
complexation ensures that, at the contrary of what happens in the gas
phase, the freshly formed complex is born cold, without the energy
needed to surmount any barrier to isomerization toward the global
energy minimum. It is remarkable that this process is repeated until
the linear chain spans the whole cluster diameter, as the data of
Nauta and Miller show.  Observed spectra for CH$_3$OH and CH$_3$CN
clusters~\cite{Behrens97}, for H$_2$O clusters~\cite{Nauta00a} and for
Ar$_n$+HF clusters (described in their article in this
issue~\cite{Nauta01_here}) have been interpreted as arising from the
isomers which are formed with the smallest amount of nuclear
rearrangement.

So far this self-assembling technique has been applied using stable
molecules.  Efforts are in progress in our laboratory to use CN
radicals for preparing isocyanogen-like chains or, if the presence of
a recombination barrier is detected, at least van der Waals clusters
of CN radicals. The potential applications of this technique for the
synthesis of exotic or high energy density materials are quite
evident.

\begin{figure}[bp]
\includegraphics[width=3.375 in]{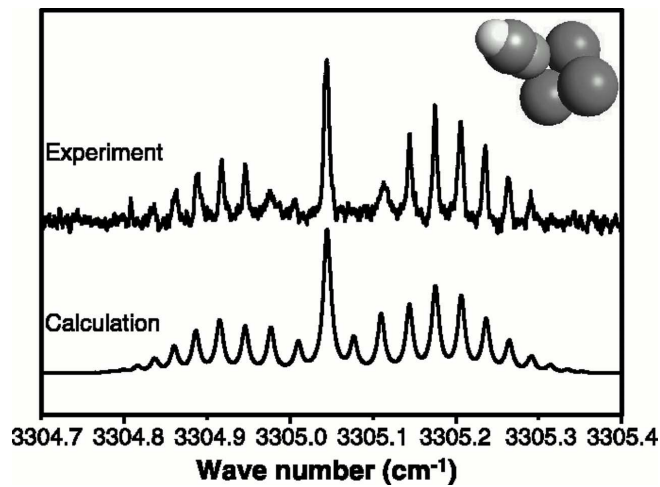}
  \caption{Rotationally resolved spectrum of the van der Waals complex
    Mg$_3$--HCN\@. The structural information extracted from this
    spectrum allows one to deduce a contraction of the
    HCN--metal-cluster bond, when going from Mg$_2$ to Mg$_3$. This in
    turn indicates the presence of strong nonadditivity effects. From
    Ref.~\onlinecite{Nauta01a}.}
  \label{fig:HCNMg3}
\end{figure}

A different, but equally useful, non-equilibrium synthesis has been
demonstrated by Nauta and Miller in an experiment~\cite{Nauta01a} in
which a cluster of reactive metal atoms (Mg) has been attached to an
IR active molecule (HCN), and the rotational spectrum of the resulting
complex has been resolved (Figure~\ref{fig:HCNMg3}).  While clusters
of metal atoms are bound by van der Waals or metallic forces according
to their size~\cite{Busani98,Bowen01,Diederich01,Kohn01}, they are in
both cases very difficult to study because their UV spectra are made
of broad lines and their IR spectra, when present, appear at rather
low frequencies, where lasers are not yet available.  Probing the IR
spectrum of a molecule attached to the atomic cluster yields valuable
information on the latter via an analysis of the rotationally resolved
spectrum or via the spectral shifts of the molecule's vibrations.
While the information contained in rotationally resolved spectra is at
present clouded by our imperfect knowledge of the helium contributions
(see Section~\ref{sec:rotation}), the type of spectrum (i.e., the
symmetries of complexes) and shifts are already now sources of
reliable information. We anticipate that this type of experiments will
provide a wealth of information on metal-ligand interactions and on
the nature of bonding in chemisorbed systems.

Most large organic molecule, including those of biological relevance,
have a number of isomers populated at room temperature, particularly
arising from rotation about single bonds (conformers). When these
molecules are cooled to low temperature, either in MI or supersonic
jet spectroscopies, isomers separated by high barriers do not
interconvert, while those separated from lower-lying minima by low
barriers will cool out.  Multiple isomers have been observed by HENDI
spectroscopy as well.  As an example, see the study of glycine by
Huisken \etal~\cite{Huisken99}, where the same three conformers were
observed in He nanodroplets and in Ne and Ar matrices.  In general,
however, the different cooling rates should result in different
distributions of conformers in the different spectroscopies.
Spectra of tryptophan and tyrosine in helium have been
interpreted in terms of suppression of certain conformers in helium
compared to
those observed in seeded beam expansions~\cite{Lindinger99}.  The
more rapid quenching, lower final temperatures, and shorter interaction
times of HENDI suggest that the conformer distribution observed by
this method should more closely resemble that of the initial ensemble.
The much smaller spectral shifts in helium should make conformer
assignments based upon quantum chemistry calculations significantly
more reliable than for MI\@.  The use of pendular spectroscopy in
helium, which allows measurement of the permanent dipole moments and
the angle between that and the vibrational transition moment, should
further greatly enhance reliability of assignments.

\subsection{Large-amplitude molecular motions in liquid helium}
\label{sec:nh3}

One of the great differences between electronic and ro-vibrational
spectroscopy in He droplets is the extent to which helium is displaced
upon excitation of the molecule. Helium displacement arises from the
repulsive interaction (which can be viewed in terms of the Pauli
exclusion principle) between the electron clouds of He atoms and that
of the molecule. Electronic excitation results in general in an
expansion of the electron cloud, with an energetic cost that is
responsible for the experimentally observed blue shifts of electronic
spectra of atomic and molecular species immersed in liquid He.
Conversely, the volume and shape of the electron cloud are not
significantly changed upon ro-vibrational excitation. The observed
band shifts are small, mostly towards the red, and are determined by a
delicate balance between the attractive and repulsive forces in the
ground and excited state~\cite{Callegari00c} (see
Section~\ref{sec:shifts}).  As a first approximation, one can say that
molecular stretch vibrations, with few exceptions the only modes
studied so far, are too stiff to be significantly affected by the
presence of the helium.

This statement does not hold true for large-amplitude modes which are,
somewhat by definition, softer (the FWHM of the 1-D harmonic
oscillator is proportional to $(m\omega)^{-1/2}$ or $(k m)^{-1/4}$).
They are particularly relevant not only because they displace more
helium (supposedly coupling better to He bulk modes), but also because
they are usually bending modes. Tunneling barriers commonly occur for
such modes, and are expected to be increased by the presence of the
helium. For the above reasons, these mode are a useful probe of the
molecule-helium interaction.

The prototypical transition is the \vi{2} umbrella motion of
NH$_{3}$~\cite{Behrens98}: the hindrance induced by the helium causes
a blue shift of 17.4~\wn, an order of magnitude larger than what is
commonly observed. Within the limits imposed by the coarse resolution
of the spectrum \mbox{($\sim 1$~\wn)}, the excited-state inversion
splitting is found to be decreased by 31\%. At the time, only
speculations could be made as for the cause: it was suggested that
helium might increase either the tunneling barrier and/or the effective
mass of the oscillator, or even change the symmetry of the inversion
potential, thus suppressing tunneling overall. With the knowledge we
have gained in the meantime from the study of molecular rotation, we
can reasonably say that the first two are going to be important while
the third is probably not. Similar observations have been made for the
\vi{1} and \vi{2} modes of the ammonia dimer~\cite{Behrens97}. In
this system the inversion tunneling is already suppressed by the
partner molecule, and the presence of the helium has, if any, a
smaller effect. For the same reason, the large blue shift observed for
the monomer is not present here. Interchange tunneling is instead
affected by the presence of the helium: the splitting is reduced from
5--10~\wn\ to 2~\wn.

A capital step forward is the study of the HF dimer by Nauta and
Miller~\cite{Nauta00b}: the use of a low power continuously tunable
laser and the added diagnostics provided by Stark fields allowed a
truly quantitative study of tunneling motions.  The interchange
tunneling splitting, like in the ammonia dimer, is observed to be
decreased, by an amount that is accurately quantified (-41\%).  Nauta
and Miller propose that this is due to an increased barrier; the
barrier increase is explained with the observation that the transition
state is nonpolar, hence has less energy gain from polarization of the
helium than the polar equilibrium states.  While it is
difficult to rule out additional contributions, such as those proposed
for ammonia, this one must definitely be present, since the simple
model used in Ref.~\onlinecite{Nauta99b} shows quantitative agreement
with the experiment. Nauta and Miller observe that this interpretation
also explains a rather unusual observation. The $A$ rotational
constant is \textit{increased} (from 31.96 to 33.3~\wn) upon solvation
in helium, a fact which is indicative of a more linear structure. Such
a structure exhibits a larger dipole moment, and is therefore
consistent with a polarization energy gain.
% Further support to an increased tunneling barrier comes from the
% observation that an electric field significantly smaller than in the
% gas phase (\mbox{$\sim 50$~kV/cm} vs \mbox{$\sim 160$}) is necessary
% to quench the tunnelling motion.

The experiment also measured the \vi{2}+\vi{5} and \vi{1}+\vi{4}
bands, from which the frequencies of the soft modes involving the van
der Waals bond (\vi{4}, intermolecular stretch, and \vi{5},
intermolecular bend) are reconstructed. Both are found to be
increased, relative to the gas phase, by \mbox{$\approx 4$\%}
(\mbox{$\sim 133$~\wn} up from 127.57, and \mbox{$\sim 185$~\wn} up
from 178.67, respectively), which is a huge value compared to any
other mode observed so far. This observation is in line with the
general expectation that soft modes will be more strongly affected by
the hindering presence of the helium.

\subsection{$^{3}$He and mixed $^{4}$He/$^{3}$He droplets}
\label{sec:3He}
At present, rotationally resolved infrared spectra in pure or mixed
$^{3}$He droplets have been only obtained for of
SF$_{6}$~\cite{Harms97b,Harms99a} and
OCS~\cite{Grebenev98,Grebenev00a}. In mixed droplets the slightly
lower zero point energy of $^{4}$He atoms favors their localization
around the dopant molecule\cite{Pi99}. Thus a small doped $^{4}$He
cluster within the large $^{3}$He droplet is formed; clusters with a
small number of $^{4}$He atoms ($N_{4}=20$--30) produce considerably
broadened spectra (ascribed to inhomogeneous broadening induced by the
spread of N$_{4}$).  Otherwise the resulting spectra are, to first
approximation, the same as in pure $^{4}$He, except the temperature is
lower (0.15~K), being dictated by evaporation of $^{3}$He.  As the
concentration of $^{4}$He in the expanding gas mixture is increased
from 0.1 to 4\% the temperature rises gradually to the limit value of
pure $^{4}$He droplets, 0.38~K~\cite{Harms99a}.

Because of the lower number density of $^{3}$He, the matrix shift of
the band origin (Sec.~\ref{sec:shifts}) is also affected. The band
origin shift of SF$_{6}$ has been used to estimate the number of
$^{4}$He atoms in the inner cluster\cite{Harms99a}; small shifts
within the R branch suggest that the rotational constant of SF$_{6}$
in mixed droplets with a large number of $^{4}$He atoms (200--500) is
$\approx 3\%$ larger then in pure $^{4}$He droplets. The effect is
more evident for OCS, where the rotational constant drops by up to
10\%, at $N_{4}=60$.\cite{grebenev_thesis,Grebenev00d}

Studies of isotopically pure $^{3}$He droplets were limited to OCS
which has a simpler Hamiltonian and a better resolved
spectrum\cite{Grebenev98,Grebenev00a}.  Whereas in pure $^{4}$He the
spectrum shows the well resolved rotational structure expected for a
linear molecule, in pure $^{3}$He droplets the spectrum collapses into
a broad single band, very similar to what is expected in a classical
fluid.  Since the binary interaction potentials of OCS with $^{3}$He
and $^{4}$He atoms are virtually identical, the difference must be due
to the different states of the droplet: superfluid for $^{4}$He and
normal fluid for $^{3}$He.  Molecules in non-superfluid $^{3}$He are
rapidly exchanging their energy and angular momentum with the medium
as compared to superfluid $^{4}$He\cite{Babichenko99}. Thus the
ability of the dopant molecule to rotate coherently over many periods
is a microscopic manifestation of superfluidity.  Recently, the
spectrum in pure $^{3}$He has also been fitted to an unresolved
rotational structure; this gives a rotational constant 1.6 times
smaller than in $^{4}$He.\cite{Grebenev00d}

The onset of superfluidity in small $^{4}$He clusters has been
investigated by intentionally doping isotopically pure $^{3}$He
droplet containing one OCS molecule with a controlled number (up to
100) of $^{4}$He atoms.\cite{Grebenev98} The gradual appearance of a
sharp rotational structure indicates that the onset of superfluidity
is a smooth process, and is almost complete in cluster of just
60$^{4}$He atoms, corresponding to approximately two solvation layers.
While it is understood that preservation of rotational coherence
depends on lack of accessible excitation of the helium, in turn linked
to the Bose permutation symmetry of the first solvation
shell(s)~\cite{Babichenko99,Lehmann01}, the detailed mechanism of this
effect is still to be understood.

Further addition of the $^{4}$He atoms do not lead to qualitative
change of the spectrum.  The appearance of particularly sharp lines in
the P branch ($< 100$~MHz) has been attributed to cancellation of
rotational and vibrational contributions to inhomogeneous
broadening.\cite{grebenev_thesis}

\subsection{Molecule/hydrogen clusters in helium droplets}
\label{sec:hydrogen}

The HENDI study of van der Waals clusters containing various
isotopomers of molecular hydrogen deserves special attention because
of the highly quantum nature of the latter molecule. Just like the
magnesium clusters discussed in Section~\ref{sec:noneq}~\cite{Nauta01a},
the arrangement and properties of hydrogen molecules are studied
through their effect on the infrared spectrum of a larger molecule
such as OCS~\cite{Grebenev01,Grebenev01b,Grebenev00c}.

The first study of an \mbox{OCS--H$_2$} (--HD, --D$_2$) van der Waals
complex~\cite{Grebenev01} has indicated that the hydrogen molecule
lies in the ring-shaped locus of the potential minimum around OCS,
with a rather large delocalization out of the average
\mbox{OCS--hydrogen} plane. It is not clear, however, to what degree
the use of Kraitchman's equations (which are used to derive the
position of an isotopically substituted atom from the resulting change
in the moments of inertia, assuming a rigid structure~\cite{Gordy84})
is justified given the large amplitude of the hydrogen vibration.
However, the resulting \mbox{OCS--H$_2$} structure was found to be in
reasonable agreement with the predictions of recent \emph{ab initio}
calculations~\cite{Higgins99}.

The same authors have found~\cite{Grebenev01b} that upon
addition of more \emph{para}-hydrogen molecules to the
\mbox{OCS--(\pH)$_n$} van der Waals complex in pure $^4$He (0.38~K)
and mixed $^4$He/$^3$He (0.15~K) droplets, the ro-vibrational spectrum
rapidly approaches that of a symmetric top, until, for $n=5$ and 6,
the Q-branch disappears and the symmetry of the spectrum is
that of a linear molecule again, just as for a pure OCS molecule in a
helium droplet.  This is most probably due to the fact that the
bosonic ($I=0$) \pH\ molecules form a complete ring of C$_{n v}$
($n=5$--6) symmetry around OCS\@. As discussed for helium excitations in
Section~\ref{sec:openquestions}, azimuthal excitations of such a ring
are forbidden if $K$ is not an integer multiple of $n$. In the cold
environment of the helium droplets, such high-$K$ levels are not
populated, and there can be no Q-branch.

In the same article, a very interesting comparison has been made with
the same complex but using \emph{ortho}-deuterium (\oD) instead of
\pH. One would expect that the 6-fold nuclear spin degeneracy of \oD\
makes the probability for building a ring of indistinguishable \oD\
molecules equal to $6^{1-n}$, which amounts to 0.8\% for $n=5$ and
0.1\% for $n=6$. The resulting decreases in intensity of the
Q-branches of \mbox{OCS--(\oD)$_n$} complexes are therefore
negligible.  Indeed, the spectra of all \mbox{OCS--(\oD)$_n$} clusters
up to $n=8$ feature a Q-branch.

Most surprisingly, the same authors have found~\cite{Grebenev00c} that
the Q-branch disappears again for \mbox{OCS--(\pH)$_n$} clusters with
$n=14$--16 at 0.15~K, but not at 0.38~K\@. For these values of $n$,
the \pH\ molecules probably form several rings around OCS, with
different numbers $n_i$ of molecules per ring.  One could imagine a
situation where two rings are formed, with $n_1=6$ and $n_2=8$ \pH\ 
molecules, with a C$_{2 v}$ structure.  Based upon the results for the
OCS--H$_2$ monomer, a moment of inertia of $I_a \approx 370$~u \AA$^2$
can be estimated for this structure.  This would imply that the lowest
accessible state with rotational angular moment along the OCS axis
($K=2$) will have an excitation energy of 0.26~K, allowing significant
population at 0.38~K but not at 0.15~K\@.  The next two \pH\ molecules
could then be added at both ends of the OCS on the molecular axis,
preserving the C$_{2 v}$ structure, and finally resulting in a filled
shell of 16 \pH\ molecules around the OCS (compare to a filled $^4$He
shell that consists of 17 molecules~\cite{Kwon00}).  This picture
neglects any additional reduction of the rotational constant by the
helium, which is expected to be small because such a highly symmetric
structure may not induce much angular anisotropy in the helium.  The
authors of Ref.~\onlinecite{Grebenev00c} propose the radically
different explanation that at the lower temperature (0.15~K), large
delocalization of the \pH\ molecules leads to bosonic exchange and
thus to a ``superfluid'' state of the hydrogen rings, which
effectively makes the symmetry group C$_{\infty v}$ and forbids any
states with $K \neq 0$. At higher temperature (0.38~K), however,
delocalization is less, and thermal population of $|K|>0$ states
yields a Q-branch.  The field of (\pH)$_n$ clusters will need further
study in order to corroborate this claim of superfluidity.

\section{Conclusion}

HENDI spectroscopy has evolved from earlier molecular beam techniques
for isolated and complexed atoms and molecules and has, in a rather
short time, reached the point where it can be applied to the study of
a wide variety of problems.  Ro-vibrational spectroscopy in this novel
extremely cold and finite environment reveals rich spectra, often with
much more detail than in other condensed phases.  The field has made
significant progress in developing the experimental and theoretical
tools needed to extract relevant information from these spectra.
However, a number of fundamental questions remain which principally
relate to the nature of the interactions of the molecular motions with
the highly quantum many-body dynamics of the helium.  We still do not
understand the mechanism by which molecules come into equilibrium with
the droplets, nor do we know precisely the time scales for many of the
relaxation processes that occur.

Despite the fact that much work remains to be done on these basic
issues, the field is also rapidly making the transition from using
molecules to study the physics of the droplets to using droplets to
create and study novel chemical species.  Vibrational spectroscopy has
produced the bulk of the results in this direction, due to the high
resolution available and the nearly additive effects from multiple
perturbers.  New tools, especially continuously tunable IR lasers at
longer wavelengths than $3.5$~$\mu$m and the development of sources of
radicals for pickup by the droplets, promise to greatly expand the
range of Chemical Physics that can be done with the smallest and
coldest chemical vials yet devised.

\acknowledgments

The authors would like to thank Udo Buch, David Farrelly, Martina
Havenith, and Andrej Vilesov for carefully reading and commenting on
this manuscript. This work was supported by grants from the National
Science Foundation and the Air Force Office of Scientific Research.

\bibliographystyle{prsty}
%\bibliography{bielefeld,bochum,buck,he_fog,MPI,Nijmegen,Princeton,UNC,UIUC,review,rev98,extra}
\bibliography{920141JCP}

\begin{thebibliography}{100}

\bibitem{Toennies98}
J.~P. Toennies and A.~F. Vilesov, Annu.\ Rev.\ Phys.\ Chem. {\bf 49},  1
  (1998).

\bibitem{Whaley98}
K.~B. Whaley,  in {\em Advances in Molecular Vibrations and Collision
  Dynamics}, edited by J.~M. Bowman and Z. Bacic (JAI Press, Inc., Greenwich,
  Conn., 1998), Vol.~III, p.\ 397.

\bibitem{Stewart83}
G.~M. Stewart and J.~D. McDonald, J.~Chem.\ Phys. {\bf 78},  3907  (1983).

\bibitem{Chang93}
H.-C. Chang and W. Klemperer, J.~Chem.\ Phys. {\bf 98},  2497  (1993).

\bibitem{Miller92}
R.~E. Miller,  in {\em Atomic and molecular beam methods}, edited by G. Scoles
  (Oxford University Press, New York, 1992), Vol.~2, Chap.~6, p.\ 192.

\bibitem{Gough77}
T.~E. Gough, R.~E. Miller, and G. Scoles, Appl.\ Phys.\ Lett. {\bf 30},  338
  (1977).

\bibitem{Nesbitt94}
D.~J. Nesbitt, Annu.\ Rev.\ Phys.\ Chem. {\bf 45},  367  (1994).

\bibitem{Klots85}
C.~E. Klots, J.~Chem.\ Phys. {\bf 83},  5854  (1985).

\bibitem{Hutson90}
J.~M. Hutson, Annu.\ Rev.\ Phys.\ Chem. {\bf 41},  123  (1990).

\bibitem{Gough85}
T.~E. Gough, M. Mengel, P.~A. Rowntree, and G. Scoles, J.~Chem.\ Phys. {\bf
  83},  4958  (1985).

\bibitem{Gough83}
T.~E. Gough, D.~G. Knight, and G. Scoles, Chem.\ Phys.\ Lett. {\bf 97},  155
  (1983).

\bibitem{Tabbert97}
B. Tabbert, H. G{\"u}nther, and G. zu~Putlitz, J.~Low Temp.\ Phys. {\bf 109},
  653  (1997).

\bibitem{Tam99}
S. Tam, M. Fajardo, H. Katsuki, H. Hoshina, T. Wakabayashi, and T. Momose,
  J.~Chem.\ Phys. {\bf 111},  4191  (1999).

\bibitem{Goyal92b}
S. Goyal, D.~L. Schutt, and G. Scoles, Phys.\ Rev.\ Lett. {\bf 69},  933
  (1992).

\bibitem{Grebenev98}
S. Grebenev, J. Toennies, and A. Vilesov, Science {\bf 279},  2083  (1998).

\bibitem{Lehmann98}
K.~K. Lehmann and G. Scoles, Science {\bf 279},  2065  (1998).

\bibitem{Northby01_here}
J. Northby, J.~Chem.\ Phys. {\bf xx},  xxxx  (2001).

\bibitem{cryo}
e.g., Advanced Research Systems; Sumitomo Heavy Industries Ltd.; Helix
  Technology, 2001.

\bibitem{Buchenau90}
H. Buchenau, E.~L. Knuth, J. Northby, J.~P. Toennies, and C. Winkler, J.~Chem.\
  Phys. {\bf 92},  6875  (1990).

\bibitem{Harms98a}
J. Harms, J. Toennies, and F. Dalfovo, Phys.\ Rev.\ B {\bf 58},  3341  (1998).

\bibitem{Harms01}
J. Harms, J. Toennies, M. Barranco, and M. Pi, Phys.\ Rev.\ B {\bf 63},  4513
  (2001).

\bibitem{habe94}
H. Haberland,  in {\em Clusters of atoms and molecules: {T}heory, experiment,
  and cluster of atoms}, Vol.~52 of {\em Springer Series in Chemical Physics},
  edited by H. Haberland (Springer Verlag, New York, 1994), Chap.~3.

\bibitem{Smith98}
R.~A. Smith, T. Ditmire, and J.~W.~G. Tisch, Rev.\ Sci.\ Instr. {\bf 69},  3798
   (1998).

\bibitem{knut94}
E.~L. Knuth, B. Schilling, and J.~P. Toennies,  in {\em International Symposium
  on Rarefied Gas Dynamics (19$^{th}$ : 1994 University of Oxford)} (Oxford
  University Press, Oxford, 1995), Vol.~19, p.\ 270, $\langle$N$\rangle$ was
  calculated according to the scaling law given in this reference; since no
  explicit dependence of $\langle$N$\rangle$ on the scaling parameter $\Gamma$
  is given, the data in fig. 1 were fitted to a line obtaining:
  $\ln(\langle$N$\rangle)=2.44+2.25 \ln(\Gamma)$.

\bibitem{Harms98b}
J. Harms and J.~P. Toennies, J.~Low Temp.\ Phys. {\bf 113},  501  (1998).

\bibitem{Harms97a}
J. Harms, J.~P. Toennies, and E.~L. Knuth, J.~Chem.\ Phys. {\bf 106},  3348
  (1997).

\bibitem{Knuth99}
E.~L. Knuth and U. Henne, J.~Chem.\ Phys. {\bf 110},  2664  (1999).

\bibitem{Hartmann99}
M. Hartmann, N. Portner, B. Sartakov, J.~P. Toennies, and A.~F. Vilesov,
  J.~Chem.\ Phys. {\bf 110},  5109  (1999).

\bibitem{Ceperley95}
D.~M. Ceperley, Rev.\ Mod.\ Phys. {\bf 67},  279  (1995).

\bibitem{Abraham70}
B.~M. Abraham, Y. Eckstein, J.~B. Ketterson, M. Kuchnir, and P.~R. Roach,
  Phys.\ Rev.\ A {\bf 1},  250  (1970).

\bibitem{Jortner92}
J. Jortner, Z.\ Phys.\ D--Atoms Mol.\ Clusters {\bf 24},  247  (1992).

\bibitem{Gspann82}
J. Gspann,  in {\em {P}hysics of electronic and atomic collisions}, edited by
  S. Datz (North Holland, Amsterdam, 1982), p.\ 79.

\bibitem{Brink90}
D. Brink and S. Stringari, Z.\ Phys.\ D--Atoms Mol.\ Clusters {\bf 15},  257
  (1990).

\bibitem{Harms97b}
J. Harms, M. Hartmann, J. Toennies, A. Vilesov, and B. Sartakov, J.~Mol.\
  Spectr. {\bf 185},  204  (1997).

\bibitem{beamdynamics}
Beam Dynamics, Inc., Eden Prairie, MN, 2001.

\bibitem{Miller88}
D.~L. Miller,  in {\em Atomic and molecular beam methods}, edited by G. Scoles
  (Oxford University Press, New York, 1988), Vol.~1, Chap.~2, p.\ 17.

\bibitem{Stephens83}
P. Stephens and J. King, Phys.\ Rev.\ Lett. {\bf 51},  1538  (1983).

\bibitem{Gspann95}
J. Gspann, Z.~Phys.\ B {\bf 98},  405  (1995).

\bibitem{Farnik98}
M. Farnik, U. Henne, B. Samelin, and J. Toennies, Phys.\ Rev.\ Lett. {\bf 81},
  3892  (1998).

\bibitem{Harms99b}
J. Harms and J. Toennies, Phys.\ Rev.\ Lett. {\bf 83},  344  (1999).

\bibitem{Harms99a}
J. Harms, M. Hartmann, B. Sartakov, J. Toennies, and A. Vilesov, J.~Chem.\
  Phys. {\bf 110},  5124  (1999).

\bibitem{isotec}
Isotec, Inc. \$175/L for 99.95\% and \$280/L for Grade 6, both at STP.

\bibitem{Schollkopf94}
W. Sch\"ollkopf and J. Toennies, Science {\bf 266},  1345  (1994).

\bibitem{Guardiola00}
R. Guardiola and J. Navarro, Phys.\ Rev.\ Lett. {\bf 84},  1144  (2000).

\bibitem{Barranco97}
M. Barranco, J. Navarro, and A. Poves, Phys.\ Rev.\ Lett. {\bf 78},  4729
  (1997).

\bibitem{chin95}
S.~A. Chin and E. Krotscheck, Phys.\ Rev.\ B {\bf 52},  10405  (1995), the
  binding energy per atom ($E$, in K), as a function of droplet size $N$, is
  given by: $E=-7.21+17.71N^{-1/3}-5.95N^{-2/3}$.

\bibitem{Lewerenz95}
M. Lewerenz, B. Schilling, and J.~P. Toennies, J.~Chem.\ Phys. {\bf 102},  8191
   (1995).

\bibitem{Nauta_preprint3}
K. Nauta and R.~E. Miller, The vibrational and rotational dynamics of acetylene
  solvated in superfluid helium nanodroplets, preprint, 2001.

\bibitem{Nauta99a}
K. Nauta and R.~E. Miller, Science {\bf 283},  1895  (1999).

\bibitem{Nauta99d}
K. Nauta, D.~T. Moore, and R.~E. Miller, Faraday Discuss. {\bf 113},  261
  (1999).

\bibitem{Grebenev00c}
S. Grebenev, B. Sartakov, J. Toennies, and A. Vilesov, Science {\bf 289},  1532
   (2000).

\bibitem{Goyal93b}
S. Goyal, D.~L. Schutt, and G. Scoles, J.~Phys.\ Chem. {\bf 97},  2236  (1993).

\bibitem{Hartmann95}
M. Hartmann, R.~E. Miller, J.~P. Toennies, and A. Vilesov, Phys.\ Rev.\ Lett.
  {\bf 75},  1566  (1995).

\bibitem{Grebenev01}
S. Grebenev, B. Sartakov, J. Toennies, and A. Vilesov, J.~Chem.\ Phys. {\bf
  114},  617  (2001).

\bibitem{Grebenev01b}
S. Grebenev, E. Lugovoi, B.~G. Sartakov, J.~P. Toennies, and A.~F. Vilesov,
  Faraday Discuss. {\bf 118},  19  (2001).

\bibitem{Blume96}
D. Blume, M. Lewerenz, F. Huisken, and M. Kaloudis, J.~Chem.\ Phys. {\bf 105},
  8666  (1996).

\bibitem{Frochtenicht96}
R. Fr\"ochtenicht, M. Kaloudis, M. Koch, and F. Huisken, J.~Chem.\ Phys. {\bf
  105},  6128  (1996).

\bibitem{Huisken99}
F. Huisken, O. Werhahn, A.~Y. Ivanov, and S.~A. Krasnokutski, J.~Chem.\ Phys.
  {\bf 111},  2978  (1999).

\bibitem{Behrens97}
M. Behrens, U. Buck, R. Fr\"ochtenicht, M. Hartmann, and M. Havenith, J.~Chem.\
  Phys. {\bf 107},  7179  (1997).

\bibitem{Behrens98}
M. Behrens, U. Buck, R. Fr\"ochtenicht, M. Hartmann, F. Huisken, and F.
  Rohmund, J.~Chem.\ Phys. {\bf 109},  5914  (1998).

\bibitem{Behrens99}
M. Behrens, R. Fr\"ochtenicht, M. Hartmann, J.~G. Siebers, U. Buck, and F.~C.
  Hagemeister, J.~Chem.\ Phys. {\bf 111},  2436  (1999).

\bibitem{Kunze00}
M. Kunze, J. Reuss, J. Oomens, F. Harren, and D.~H. Parker, Z.\ Phys.\
  Chemie--Int.\ J.\ Res.\ Phys.\ Chem.\ Chem.\ Phys. {\bf 214},  1209  (2000).

\bibitem{Nauta99b}
K. Nauta and R.~E. Miller, Phys.\ Rev.\ Lett. {\bf 82},  4480  (1999).

\bibitem{Nauta00a}
K. Nauta and R.~E. Miller, Science {\bf 287},  293  (2000).

\bibitem{Nauta00b}
K. Nauta and R.~E. Miller, J.~Chem.\ Phys. {\bf 113},  10158  (2000).

\bibitem{Nauta00c}
K. Nauta and R.~E. Miller, J.~Chem.\ Phys. {\bf 113},  9466  (2000).

\bibitem{Nauta01a}
K. Nauta, D.~T. Moore, P.~L. Stiles, and R.~E. Miller, Science {\bf 292},  481
  (2001).

\bibitem{Callegari00b}
C. Callegari, I. Reinhard, K.~K. Lehmann, G. Scoles, K. Nauta, and R.~E.
  Miller, J.~Chem.\ Phys. {\bf 113},  4636  (2000).

\bibitem{Callegari00c}
C. Callegari, A. Conjusteau, I. Reinhard, K.~K. Lehmann, and G. Scoles,
  J.~Chem.\ Phys. {\bf 113},  10535  (2000).

\bibitem{even00}
U. Even, J. Jortner, D. Noy, N. Lavie, and C. Cossart-Magos, J.~Chem.\ Phys.
  {\bf 112},  8068  (2000).

\bibitem{Powers98}
P.~E. Powers, T.~J. Kulp, and S.~E. Bisson, Opt.\ Lett. {\bf 23},  159  (1998).

\bibitem{Capasso99}
F. Capasso, C. Gmachl, D.~L. Sivco, and A.~Y. Cho, Phys.\ World {\bf 12},  27
  (1999).

\bibitem{Colombelli01}
R. Colombelli, F. Capasso, C. Gmachl, A.~L. Hutchinson, D.~L. Sivco, A.
  Tredicucci, M.~C. Wanke, A.~M. Sergent, and A.~Y. Cho, Appl.\ Phys.\ Lett.
  {\bf 78},  2620  (2001).

\bibitem{Haberland94}
H. Haberland, {\em Clusters of atoms and molecules: {T}heory, experiment, and
  cluster of atoms}, Vol.~52 of {\em Springer Series in Chemical Physics}
  (Springer Verlag, New York, 1994).

\bibitem{Frochtenicht94}
R. Fr{\"o}chtenicht, J.~P. Tonnies, and A.~F. Vilesov, Chem.\ Phys.\ Lett. {\bf
  229},  1  (1994).

\bibitem{Reinhard99}
I. Reinhard, C. Callegari, A. Conjusteau, K.~K. Lehmann, and G. Scoles, Phys.\
  Rev.\ Lett. {\bf 82},  5036  (1999).

\bibitem{Conjusteau00}
A. Conjusteau, C. Callegari, I. Reinhard, K.~K. Lehmann, and G. Scoles,
  J.~Chem.\ Phys. {\bf 113},  4840  (2000).

\bibitem{Hartmann01}
M. Hartmann, A. Lindinger, J.~P. Toennies, and A.~F. Vilesov, J.~Phys.\ Chem.\
  A {\bf 105},  6369  (2001).

\bibitem{Demtroeder96}
W. Demtr{\"o}der, {\em Laser Spectroscopy}, 2nd  ed. (Springer-Verlag, Berlin,
  Heidelberg, New York, 1996).

\bibitem{Grebenev00b}
S. Grebenev, M. Havenith, F. Madeja, J.~P. Toennies, and A.~F. Vilesov,
  J.~Chem.\ Phys. {\bf 113},  9060  (2000).

\bibitem{crc70}
{\em CRC handbook of chemistry and physics}, 70$^{th}$ ed., edited by R.~C.
  Weast (CRC Press, Boca Raton, FL, 1990).

\bibitem{Reho00a}
J. Reho, U. Merker, M.~R. Radcliff, K.~K. Lehmann, and G. Scoles, J.~Chem.\
  Phys. {\bf 112},  8409  (2000).

\bibitem{Townes75}
C.~H. Townes and A.~L. Schawlow, {\em Microwave Spectroscopy} (Dover
  Publications, Inc., New York, 1975).

\bibitem{Nauta_preprint2}
K. Nauta and R.~E. Miller, Vibrational relaxation of {Ne}, {Ar}, {Kr}-{HF}
  ($v=1$) binary complexes in helium nanodroplets, preprint, 2001.

\bibitem{Loesch90}
H.~J. Loesch and A. Remscheid, J.~Chem.\ Phys. {\bf 93},  4779  (1990).

\bibitem{Friedrich91}
B. Friedrich, D.~P. Pullman, and D.~R. Herschbach, J.~Phys.\ Chem. {\bf 95},
  8118  (1991).

\bibitem{Nauta_Campargue}
K. Nauta and R.~E. Miller,  in {\em Atomic and molecular beams: {T}he state of
  the art 2000}, edited by R. Campargue (Springer Verlag, Berlin, 2001), part
  {VI}.3, p.\ 775.

\bibitem{Grebenev00a}
S. Grebenev, M. Hartmann, M. Havenith, B. Sartakov, J.~P. Toennies, and A.~F.
  Vilesov, J.~Chem.\ Phys. {\bf 112},  4485  (2000).

\bibitem{Miller_Ringberg}
R.~E. Miller, {IV} {W}orkshop on quantum fluid clusters, Ringberg Schloss,
  2000.

\bibitem{Dalfovo99}
F. Dalfovo, R. Mayol, M. Pi, and M. Barranco, Phys.\ Rev.\ Lett. {\bf 85},
  1028  (1999).

\bibitem{Lehmann99a}
K.~K. Lehmann, Mol.\ Phys. {\bf 97},  645  (1999).

\bibitem{Stienkemeier95}
F. Stienkemeier, J. Higgins, W.~E. Ernst, and G. Scoles, Phys.\ Rev.\ Lett.
  {\bf 74},  3592  (1995).

\bibitem{Hartmann96}
M. Hartmann, F. Mielke, J.~P. Toennies, A.~F. Vilesov, and G. Benedek, Phys.\
  Rev.\ Lett. {\bf 76},  4560  (1996).

\bibitem{Hu00}
C.~C. Hu, R. Petluri, and J.~A. Northby, Physica B {\bf 284},  107  (2000),
  part 1.

\bibitem{Toennies95}
J.~P. Toennies and A.~F. Vilesov, Chem.\ Phys.\ Lett. {\bf 235},  596  (1995).

\bibitem{Lehmann00}
K.~K. Lehmann, Mol.\ Phys. {\bf 98},  1991  (2000).

\bibitem{mano86}
E. Manousakis and V.~R. Pandharipande, Phys.\ Rev.\ B {\bf 33},  150  (1986).

\bibitem{ande96}
K.~H. Andersen, J. Bossy, J.~C. Cook, O.~G. Randl, and J.-L. Ragazzoni, Phys.\
  Rev.\ Lett. {\bf 77},  4043  (1996).

\bibitem{Oka73}
T. Oka, Adv.\ At.\ Mol.\ Phys. {\bf 9},  127  (1973).

\bibitem{Andronikashvili46}
E.~L. Andronikashvili, J.\ Phys.\ USSR {\bf 10},  201  (1946).

\bibitem{Kwon99b}
Y. Kwon and K.~B. Whaley, Phys.\ Rev.\ Lett. {\bf 83},  4108  (1999).

\bibitem{Ceperley_Ringberg}
D.~M. Ceperley, paper presented at the {IV} {W}orkshop on quantum fluid
  clusters, Ringberg Schloss, 2000.

\bibitem{Draeger_thesis}
E.~W. Draeger, Ph.D. thesis, University of Illinois at Urbana-Champaign, 2001,
  \url{http://archive.ncsa.uiuc.edu/Apps/CMP/draeger/draeger_thesis.ps.gz}.

\bibitem{Lamb1895}
H. Lamb, {\em Hydrodynamics}, 4 ed. ({C}ambridge {U}niversity {P}ress,
  Cambridge, 1916).

\bibitem{Callegari99}
C. Callegari, A. Conjusteau, I. Reinhard, K.~K. Lehmann, G. Scoles, and F.
  Dalfovo, Phys.\ Rev.\ Lett. {\bf 83},  5058  (1999).

\bibitem{Lee00}
E. Lee, D. Farrelly, and K.~B. Whaley, Phys.\ Rev.\ Lett. {\bf 83},  3812
  (1999).

\bibitem{Kwon00}
Y. Kwon, P. Huang, M.~V. Patel, D. Blume, and K.~B. Whaley, J.~Chem.\ Phys.
  {\bf 113},  6469  (2000).

\bibitem{Milne96}
L.~M. Milne-Thompson, {\em Theoretical hydrodynamics}, 5 ed. (Dover, New York,
  1996).

\bibitem{OrsayParis}
J. Dupont-Roc, M. Himbert, N. Pavloff, and J. Treiner, J.~Low Temp.\ Phys. {\bf
  81},  31  (1990).

\bibitem{OrsayTrento}
F. Dalfovo, A. Lastri, L. Pricaupenko, S. Stringari, and J. Treiner, Phys.\
  Rev.\ B {\bf 52},  1193  (1995).

\bibitem{dalf94}
F. Dalfovo, Z.\ Phys.\ D--Atoms Mol.\ Clusters {\bf 29},  61  (1994).

\bibitem{Lehmann00_unpublished}
K.~K. Lehmann, unpublished results, 2000.

\bibitem{Madelung26}
E. Madelung, Z.~Phys. {\bf 40},  332  (1926).

\bibitem{deBroglie26}
L. {de Broglie}, Compt.\ Rend. {\bf 183},  447  (1926).

\bibitem{Ortiz95}
G. Ortiz and D.~M. Ceperley, Phys.\ Rev.\ Lett. {\bf 75},  4642  (1995).

\bibitem{Lehmann02a}
K.~K. Lehmann and C. Callegari, Quantum Hydrodynamic Model for the enhanced
  moments of Inertia of molecules in Helium Nanodroplets: Application to
  SF$_6$, preprint, 2001, \url{http://xxx.lanl.gov/abs/physics/0109009}.

\bibitem{Blume99}
D. Blume, M. Mladenovi\'c, M. Lewerenz, and K.~B. Whaley, J.~Chem.\ Phys. {\bf
  110},  5789  (1999).

\bibitem{Lehmann01}
K. Lehmann, J.~Chem.\ Phys. {\bf 114},  4643  (2001).

\bibitem{maki74}
A.~G. Maki, J.\ Phys.\ Chem.\ Ref.\ Data {\bf 3},  221  (1974).

\bibitem{huts91}
J.~M. Hutson,  in {\em Advances in molecular vibrations and collision
  dynamics}, edited by J.~M. Bowman (JAI Press, Inc., "Greenwich, Conn.",
  1991), Vol.~1A, p.\ 1.

\bibitem{Leggett73}
A.~J. Leggett, Physica Fennica {\bf 8},  125  (1973).

\bibitem{Szalewicz01}
K. Szalewicz, Faraday Discuss. {\bf 118},  121  (2001).

\bibitem{Huisken97}
F. Huisken, M. Kaloudis, and A.~A. Vigasin, Chem.\ Phys.\ Lett. {\bf 269},  235
   (1997).

\bibitem{Eichenauer88}
D. Eichenauer and R.~J. LeRoy, J.~Chem.\ Phys. {\bf 88},  2898  (1988).

\bibitem{Kwon96}
Y. Kwon, D.~M. Ceperley, and K.~B. Whaley, J.~Chem.\ Phys. {\bf 104},  2341
  (1996).

\bibitem{Douketis82}
C. Douketis, G. Scoles, S. Marchetti, M. Zen, and A.~J. Thakkar, J.~Chem.\
  Phys. {\bf 76},  3057  (1982).

\bibitem{Hartmann96a}
M. Hartmann, F. Mielke, J.~P. Toennies, A.~F. Vilesov, and G. Benedek, Phys.\
  Rev.\ Lett. {\bf 76},  4560  (1996).

\bibitem{Nauta99c}
K. Nauta and R.~E. Miller, J.~Chem.\ Phys. {\bf 111},  3426  (1999).

\bibitem{Stienkemeier01_here}
F. Stienkemeier and A.~F. Vilesov, J.~Chem.\ Phys. {\bf xx},  xxxx  (2001).

\bibitem{Toennies01}
J.~P. Toennies, A.~F. Vilesov, and K.~B. Whaley, Phys.\ Today {\bf 54},  31
  (2001).

\bibitem{Higgins96}
J. Higgins, C. Callegari, J. Reho, F. Stienkemeier, W.~E. Ernst, K.~K. Lehmann,
  M. Gutowski, and G. Scoles, Science {\bf 273},  629  (1996).

\bibitem{Higgins00}
J. Higgins, T. Hollebeek, J. Reho, T.-S. Ho, K.~K. Lehmann, H. Rabitz, and G.
  Scoles, J.~Chem.\ Phys. {\bf 112},  1  (2000).

\bibitem{Wilks87}
J. Wilks and D.~S. Betts, {\em An introduction to liquid helium}, 2 ed.
  ({O}xford {U}niversity press, New York, 1987).

\bibitem{Nauta01_here}
K. Nauta and R.~E. Miller, J.~Chem.\ Phys. {\bf xx},  xxxx  (2001).

\bibitem{Busani98}
R. Busani, M. Folkers, and O. Cheshnovsky, Phys.\ Rev.\ Lett. {\bf 81},  3836
  (1998).

\bibitem{Bowen01}
K. Bowen, unpublished results on {M}g$_{n}^{-}$ photoelectron spectroscopy,
  2001.

\bibitem{Diederich01}
T. Diederich, T. D\"oppner, J. Braune, J. Tiggesb\"aumer, and K.-H.
  Meiwes-Broer, Phys.\ Rev.\ Lett. {\bf 86},  4807  (2001).

\bibitem{Kohn01}
A. Kohn, F. Weigend, and R. Ahlrichs, Phys.\ Chem.\ Chem.\ Phys. {\bf 3},  711
  (2001).

\bibitem{Lindinger99}
A. Lindinger, J.~P. Toennies, and A.~F. Vilesov, J.~Chem.\ Phys. {\bf 110},
  1429  (1999).

\bibitem{Pi99}
M. Pi, R. Mayol, and M. Barranco, Phys.\ Rev.\ Lett. {\bf 82},  3093  (1999).

\bibitem{grebenev_thesis}
S. Grebenev, Ph.D. thesis, Universit\"at G\"ottingen, G\"ottingen, Germany,
  2000.

\bibitem{Grebenev00d}
S. Grebenev, M. Hartmann, A. Lindinger, N. P\"ortner, B. Sartakov, J. Toennies,
  and A. Vilesov, Physica B {\bf 280},  65  (2000).

\bibitem{Babichenko99}
V.~S. Babichenko and Y. Kagan, Phys.\ Rev.\ Lett. {\bf 83},  3458  (1999).

\bibitem{Gordy84}
W. Gordy and R.~L. Cook, {\em Microwave molecular spectra} (John Wiley \& Sons,
  New York, 1984).

\bibitem{Higgins99}
K. Higgins, as reported in Ref.~\onlinecite{Grebenev01}, 1999.

\bibitem{Nauta_HCN3}
K. Nauta and R.~E. Miller, personal communication, 2001.

\bibitem{Madeja_preprint}
F. Madeja, P. Marcwick, M. Havenith, K. Nauta, and R.~E. Miller, Rotationally
  resolved infrared spectroscopy of {h}$_{2}$- and {d}$_{1}$-formic acid
  monomer in liquid {H}e-droplets, preprint, 2001.

\bibitem{Nauta_preprint4}
K. Nauta and R.~E. Miller, Rotational and vibrational dynamics of {CO}$_{2}$
  and {N}$_{2}${O} in helium nanodroplets, preprint, 2001.

\end{thebibliography}

\end{document}